\documentclass[aps,prd,reprint,amsmath,amssymb,floatfix,superscriptaddress]{revtex4-2}

\usepackage{graphicx}
\newcommand{\tabrr}[1]{{\raggedright #1\par}}
\usepackage{bm}
\usepackage{xcolor}
\usepackage[colorlinks=true,citecolor=blue,urlcolor=blue,linkcolor=blue]{hyperref}

\newcommand{\AdS}{\mathrm{AdS}}
\newcommand{\GN}{G}
\newcommand{\Geff}{\mathcal{G}}
\newcommand{\bG}{\beta_G}
\newcommand{\rhoE}{\rho_{E}}
\newcommand{\dd}{\mathrm{d}}
\newcommand{\ncol}{\tilde n_{\mathrm{col}}}
\newcommand{\Srel}{S_{\mathrm{rel}}}
\newcommand{\bom}{\bm{\omega}}
\newcommand{\by}{\bm{y}}

\begin{document}

\title{Gross--Pitaevskii--Poisson condensates from anti--de~Sitter holography}

\author{Bao-Duy Le}
\email{24130012@student.hcmus.edu.vn}
\affiliation{Faculty of Physics and Engineering Physics, University of Science, Ho Chi Minh City, Vietnam}
\affiliation{Vietnam National University, Ho Chi Minh City, Vietnam}

\date{\today}

\begin{abstract}
The Gross--Pitaevskii--Poisson (GPP) equation for a trapped, self-gravitating condensate follows from many-body theory, from Hartree theory, or as the nonrelativistic limit of Einstein--Klein--Gordon theory. We derive it holographically and show that this route does more than reproduce it: the condensate acquires a dual quantum state whose entanglement images it in three dimensions and signals its collapse. A self-interacting, charged scalar in global $\AdS_4$ is trapped by the curvature, and in the double-scaling limit $mL\to\infty$ the Einstein--Klein--Gordon--Maxwell system reduces to the trapped GPP equation, with trap frequency $1/L$, Coulomb screening of self-gravity, $\GN m^2\to\GN m^2-q^2/4\pi$, and corrections of order $(mL)^{-1}$ that we verify on back-reacted stars. Collapse is a saddle-node fold in both theories, $\omega_B\propto(N_c-N)^{1/4}$, whose fold line joins the trapped-attractive, Chavanis--Delfini and Kaup limits; at the fold the entanglement susceptibility of the dual state diverges as $(N_c-N)^{-1/2}$. In Klein coordinates the cap entanglement entropies are a Euclidean Radon transform of the bulk energy density, and an explicit inversion of kinematic-space type, which removes the boundary stress tensor at every interior point, recovers the energy density at every depth. From leading-order cap entropies we reconstruct a rotating, self-gravitating vortex lattice with an on-cloud relative $L^2$ error below $0.6\%$ for $mL\ge64$ and all seven interior vortices in place. Exact Ryu--Takayanagi surfaces of back-reacted stars show that the leading-order entropies err by $\simeq\frac32(\GN M/\langle r\rangle)\ln(L/r_{99})$, a second-order effect of the exterior field, and the image by four to five times that. When collapse is relativistic, the Jacobi instability of the Ryu--Takayanagi disk precedes the dynamical instability of the star.
\end{abstract}

\maketitle

% ======================================================================
\section{Introduction}
\label{sec:intro}

Bose--Einstein condensates bound by their own gravity appear in two disconnected settings. In astrophysics they model boson stars~\cite{Kaup1968,RuffiniBonazzola1969,Colpi1986,SchunckMielke2003,LieblingPalenzuela2017} and ultralight dark-matter halos~\cite{Hui2017}. In the laboratory, laser-induced $1/r$ interactions mimic the same Newtonian attraction~\cite{ODell2000}. In the nonrelativistic regime both are governed by the Gross--Pitaevskii--Poisson (GPP) equation, whose mass--radius relation with contact interactions was worked out by Chavanis and Delfini~\cite{Chavanis2011,ChavanisDelfini2011}.

\begin{table*}[t]
\caption{Routes to the GPP equation. The last row is the route of this paper; Ref.~\cite{BizonEvninFicek2018} took it for $\lambda=\gamma=q=0$.}
\label{tab:routes}
\begin{ruledtabular}
\begin{tabular}{p{4.0cm}p{2.1cm}p{2.2cm}p{3.2cm}p{4.0cm}}
route & small parameter & trap & long-range coupling & what comes with it \\ \hline
many-body limit~\cite{LiebSeiringerYngvason2000,LiebThirring1984,LewinNamRougerie2014,ElgartSchlein2007} & diluteness, $1/N$ & external & Newtonian pair potential & rigorous bounds on energy and dynamics \\
Hartree ansatz~\cite{Gross1961,Pitaevskii1961,RuffiniBonazzola1969} & none & external & Newtonian & --- \\
$1/c$ expansion of Einstein--Klein--Gordon~\cite{GiuliniGrossardt2012} & $1/c^2$ & none & Einstein & relativistic completion \\
AdS/CFT (this work) & $1/mL$ & geometry, $\omega_t=1/L$ & Einstein--Maxwell, $\GN m^2-q^2/4\pi$ & dual CFT state: scaling dimensions, entanglement \\
\end{tabular}
\end{ruledtabular}
\end{table*}

The GPP equation can be reached along several routes, each with its own small parameter (Table~\ref{tab:routes}). In the many-body route, the ground-state energy of trapped bosons converges to the Gross--Pitaevskii functional in the dilute limit~\cite{LiebSeiringerYngvason2000}, and that of gravitating bosons to Hartree theory in the mean-field limit~\cite{LiebThirring1984,LewinNamRougerie2014}; the mean-field dynamics of boson stars is controlled in the same way~\cite{ElgartSchlein2007}. The Hartree route postulates a single condensate mode from the outset, as in the work of Gross and Pitaevskii~\cite{Gross1961,Pitaevskii1961} and in the first boson-star models~\cite{RuffiniBonazzola1969}. In the field-theory route the GPP equation is the $1/c$ expansion of the Einstein--Klein--Gordon system~\cite{GiuliniGrossardt2012}. Along all three, a trap, when present, is put in by hand, and the equation describes the condensate and nothing else.

Anti--de~Sitter (AdS) space opens a fourth route, on which the trap is not put in by hand. The normal modes of a free scalar in global $\AdS_4$, $\omega L=\Delta+2n+\ell$, are those of an isotropic oscillator shifted by the rest energy, and Bizo\'n, Evnin and Ficek showed that the $c\to\infty$ limit of the Einstein--massive-scalar system with $\Lambda c^2$ held fixed is the Schr\"odinger--Newton--Hooke equation~\cite{BizonEvninFicek2018,Ficek2021}. Gravitating solitons in global AdS (boson stars, charged boson stars and hairy black holes) have been constructed in the holographic context~\cite{AstefaneseiRadu2003,Basu2010,Dias2012,BuchelLieblingLehner2013}, where they are dual to finite-charge states of a strongly coupled conformal field theory (CFT) on $S^2\times\mathbb R$~\cite{Maldacena1998}. This route does more than reproduce the equation. It fixes the trap frequency by the cosmological constant, makes every correction $O(1/mL)$, and gives the condensate a dual quantum state whose energy, charge and entanglement are boundary observables.

We follow the holographic route for a self-interacting, charged scalar and use the dual description to ask three questions. Is the GPP limit quantitatively controlled for the full back-reacted, self-interacting and charged system, and at what rate do the relativistic solutions approach it? How is collapse, which is Newtonian for attractive contact interactions~\cite{Ruprecht1995,ChavanisDelfini2011} and relativistic for repulsive ones~\cite{Colpi1986}, organized across the two regimes, and what does the dual state register at the instability? Which boundary quantity sees the condensate itself, rather than only its total energy?

The last question is a case of bulk reconstruction from entanglement. Lin, Marcolli, Ooguri and Stoica showed that a second derivative of the relative entropies of boundary balls is a hyperbolic Radon transform of the bulk energy density, which can be inverted point by point, in the regime where the geometry is perturbatively close to AdS~\cite{Lin2015}. Lashkari \textit{et al.} identified the relative entropy of a ball, beyond linear order, with a positive quasilocal gravitational energy~\cite{Lashkari2016}. In kinematic space the same structure takes the form of a wave equation. For perturbations of the CFT vacuum, the entanglement entropies of balls obey a free Klein--Gordon equation on an auxiliary de Sitter space~\cite{deBoerHellerMyersNeiman2016}, bulk fields enter as integrals over Ryu--Takayanagi (RT) surfaces~\cite{RyuTakayanagi2006,deBoerHaehlHellerMyers2016}, and the wave operator applied to the entropies returns the Radon transform of the bulk sources, which is the content of the linearized Einstein equation with matter~\cite{CzechLamprouMcCandlishSully2015,CzechEtAl2017}. In that language a differential operator acting on entanglement entropies already separates the boundary stress tensor from bulk matter. Linearized metric reconstruction from entanglement entropies has been carried out numerically in $\AdS_3$ with the tensor Radon transform~\cite{CaoQiSwingleTang2020}, and an explicit inversion formula relating entropy variations of ball-shaped regions to metric perturbations has been derived~\cite{Jokela2025}. What is missing, to our knowledge, is an explicit, tested reconstruction of localized, self-gravitating matter in global $\AdS_4$ with a measured error of the leading-order formula. Extremal surfaces in compact-star geometries are known to develop several branches and to leave regions that no RT surface enters~\cite{Hubeny2012,Nogueira2013,Freivogel2015}; how the onset of this behavior compares with the dynamical instability of the star has not, to our knowledge, been examined.

Section~\ref{sec:model} fixes the bulk model. Section~\ref{sec:gpp} derives the GPP limit together with the parameter map~\eqref{eq:dict}; the Maxwell field enters as a Coulomb screening of self-gravity, $\Geff=\GN m^2-q^2/4\pi$. Section~\ref{sec:fold} treats collapse as a saddle-node fold common to both theories and assembles a single collapse diagram. Section~\ref{sec:ee} is the holographic part. The hemisphere entropy separates exactly, at linear order in $\GN$, into a modular energy and a bulk relative entropy. In Klein coordinates the cap entropies are Radon data of the energy density, and the inversion~\eqref{eq:inversion} uses the entanglement entropy alone; in the GPP regime it becomes the tomography relations~\eqref{eq:tomo} and~\eqref{eq:gppct}. Section~\ref{sec:num} describes the numerics and their validation, and Appendix~\ref{sec:droplet} treats the self-bound branch at $\lambda<0<\gamma$. Section~\ref{sec:results} collects the results, among them a three-dimensional reconstruction of a rotating vortex lattice, exact RT surfaces of back-reacted stars, and the Jacobi instability of the RT disk, which precedes relativistic collapse.

% ======================================================================
\section{Model}
\label{sec:model}

We work with $\hbar=c=1$ and $\kappa=8\pi\GN$, and consider
\begin{multline}
S=\int\dd^4x\sqrt{-g}\,\Big[\frac{R-2\Lambda}{2\kappa}-\frac14F_{\mu\nu}F^{\mu\nu}\\
-|D_\mu\Psi|^2-V(|\Psi|^2)\Big],
\label{eq:action}
\end{multline}
with $\Lambda=-3/L^2$, $D_\mu=\nabla_\mu-iqA_\mu$, and
\begin{equation}
\begin{aligned}
V(x)&=m^2x+\frac{\lambda}{2}x^2+\frac{\gamma}{3}x^3,\\
U(x)&\equiv V'(x)=m^2+\lambda x+\gamma x^2 .
\end{aligned}
\label{eq:V}
\end{equation}
Here $\lambda$ is dimensionless and $\gamma$ has dimension (length)$^2$. Stationary, spherically symmetric solutions take the form
\begin{equation}
\begin{gathered}
\dd s^2=-f e^{-\chi}\dd t^2+\frac{\dd r^2}{f}+r^2\dd\Omega_2^2,\\
\Psi=\psi(r)e^{-i\omega t},\qquad A=\phi(r)\,\dd t .
\end{gathered}
\label{eq:ansatz}
\end{equation}
With $\Omega(r)\equiv\omega+q\phi(r)$ the field equations are
\begin{align}
&\psi''+\Big(\frac{f'}{f}-\frac{\chi'}{2}+\frac2r\Big)\psi'+\Big(\frac{e^{\chi}\Omega^2}{f^2}-\frac{U(\psi^2)}{f}\Big)\psi=0, \label{eq:KG}\\
&\phi''+\Big(\frac2r+\frac{\chi'}{2}\Big)\phi'-\frac{2q\Omega\psi^2}{f}=0, \label{eq:Max}\\
&\chi'=-2\kappa r\Big(\psi'^2+\frac{e^{\chi}\Omega^2\psi^2}{f^2}\Big), \label{eq:chi}\\
&f'+\frac{f-1}{r}-\frac{3r}{L^2}+\kappa r\,\rhoE=0, \label{eq:f}
\end{align}
where the energy density is
\begin{equation}
\rhoE=f\psi'^2+\frac{e^{\chi}\Omega^2\psi^2}{f}+V(\psi^2)+\frac12e^{\chi}\phi'^2 .
\label{eq:rhoE}
\end{equation}
Regularity at the origin requires $f(0)=1$ and $\psi'(0)=\phi'(0)=0$. We impose $\chi(\infty)=0$ and $\phi(\infty)=0$, so that $\omega$ is the frequency measured on the boundary. The scalar must decay without source, $\psi\sim r^{-\Delta}$ with
\begin{equation}
\Delta=\tfrac32+\sqrt{\tfrac94+m^2L^2},
\label{eq:Delta}
\end{equation}
which for $m^2>0$ lies far above the Breitenlohner--Freedman window~\cite{BreitenlohnerFreedman1982}. Outside the matter, Eqs.~\eqref{eq:Max}--\eqref{eq:f} integrate to the Reissner--Nordstr\"om--AdS form
\begin{equation}
\begin{aligned}
f&=1+\frac{r^2}{L^2}-\frac{2\GN M}{r}+\frac{\kappa q^2N^2}{32\pi^2r^2},\\
N&=8\pi\int_0^\infty\!\dd r\,\frac{r^2e^{\chi/2}\Omega\psi^2}{f},
\end{aligned}
\label{eq:MN}
\end{equation}
with $M$ the mass and $N$ the conserved $U(1)$ charge in units of $q$. The mass function $m(r)=4\pi\int_0^r\rhoE r'^2\dd r'$ satisfies $2\GN m(r)=r(1+r^2/L^2-f)$ identically, and $M=m(\infty)$. Since $\rhoE$ includes the electrostatic energy, outside the matter $m(r)=M-q^2N^2/8\pi r$, which reproduces Eq.~\eqref{eq:MN}. Along any one-parameter family of solutions,
\begin{equation}
\dd M=\omega\,\dd N .
\label{eq:firstlaw}
\end{equation}
In the dual CFT on $S^2\times\mathbb R$, $ML$ is the scaling dimension of the charge-$N$ state that the soliton represents, and $\Delta$ is the dimension of the single-trace operator dual to $\Psi$.

% ======================================================================
\section{The Gross--Pitaevskii--Poisson limit}
\label{sec:gpp}

\subsection{Reduction}

Let $\ell\equiv(L/m)^{1/2}$ and $\epsilon\equiv(mL)^{-1}=\ell^2/L^2$. The condensate lives at $r=O(\ell)$, where $r^2/L^2=O(\epsilon)$ and the AdS curvature acts as a weak quadratic potential. Write $\omega=m+\delta\omega$ and $\Psi=e^{-imt}\varphi/\sqrt{2m}$, so that $N=\int|\varphi|^2\dd^3x$ at leading order. Expand $g_{tt}=-(1+r^2/L^2+2\Phi_N)$. To first order in $\Phi_N$, in $\delta\omega/m$, in $q\phi/m$, and in $\lambda|\Psi|^2/m^2$, Eq.~\eqref{eq:KG} becomes $\nabla^2\psi+[2m(\delta\omega+q\phi)-2m^2(r^2/2L^2+\Phi_N)-\lambda\psi^2-\gamma\psi^4]\psi=0$. Equations~\eqref{eq:Max}--\eqref{eq:f} reduce to $\nabla^2\Phi_N=4\pi\GN m|\varphi|^2$ and $\nabla^2\phi=q|\varphi|^2$. Dividing by $2m$ gives
\begin{equation}
i\partial_t\varphi=\Big[-\frac{\nabla^2}{2m}+\frac12m\omega_t^2r^2+g_2|\varphi|^2+g_3|\varphi|^4+U_{\rm LR}\Big]\varphi,
\label{eq:GPP}
\end{equation}
with
\begin{equation}
\begin{gathered}
\omega_t=\frac1L,\qquad g_2=\frac{\lambda}{4m^2},\qquad g_3=\frac{\gamma}{8m^3},\\
\nabla^2U_{\rm LR}=4\pi\Geff|\varphi|^2,
\end{gathered}
\label{eq:couplings}
\end{equation}
and the screened long-range coupling
\begin{equation}
\Geff=\GN m^2-\frac{q^2}{4\pi}.
\label{eq:Geff}
\end{equation}
The Coulomb term enters with the opposite sign to gravity because $U_{\rm LR}=m\Phi_N-q\phi$. Self-gravity therefore survives only for $q<q_*\equiv m\sqrt{4\pi\GN}$. The no-force value $q_*$ reproduces the existence bound for charged boson stars~\cite{JetzerVanDerBij1989}. Large $q$ is thus not a weak-gravity GPP regime but the probe limit~\cite{HHH2008b}, in which gravity decouples and Coulomb repulsion dominates.

The reduction is local and does not use the spherical ansatz~\eqref{eq:ansatz}, which we need only for the back-reacted stars of Sec.~\ref{sec:num}. For a general stationary configuration in the gauge $g_{tt}=-(1+r^2/L^2+2\Phi_N)$, $A=\phi\,\dd t$, the same expansion applies pointwise. A state that is stationary in a frame rotating with $\Omega_{\rm rot}$ about $\hat z$, $\Psi(t,\bm x)=e^{-i\omega t}\psi(\mathsf R_{-\Omega_{\rm rot}t}\bm x)$ with $\mathsf R_\alpha$ the rotation by $\alpha$ about $\hat z$, gives the time-independent GPP equation with the Hamiltonian replaced by $H-\Omega_{\rm rot}L_z$. The gravitomagnetic components $g_{ti}$ generated by the condensate current, and the magnetic field generated by the charge current, couple through $\bm v\cdot\bm A$ with $|\bm v|\sim\Omega_{\rm rot}\ell\le\ell/L$, so they are smaller than $\Phi_N$ and $\phi$ by $O(\epsilon)$ and drop out at leading order. A rotating frame is defined on the whole boundary sphere only for $\Omega_{\rm rot}L<1$, where its equator moves slower than light~\cite{HawkingHunterTaylorRobinson1999}; in the limit this is the trap condition $\Omega_{\rm rot}<\omega_t$.

\paragraph*{Regime of validity.}
The bulk description is classical when $L^2/\GN\gg1$, which sets the central charge of the dual CFT. With $\bG=O(1)$ and $q=0$, Eq.~\eqref{eq:dict} below gives $N=\bG\,(L^2/\GN)\,(mL)^{-5/2}$, so mean-field theory needs $L^2/\GN\gg(mL)^{5/2}$. The bulk couplings follow as $\lambda=4(\beta_2/\bG)(mL)^3\GN/L^2$ and $\gamma m^2=8(\beta_3/\bG^2)(mL)^7(\GN/L^2)^2$, so for $\beta_{2,3}=O(1)$ they are perturbative only if $L^2/\GN\gg(mL)^3$ and, when $\beta_3\neq0$, $L^2/\GN\gg(mL)^{7/2}$. The double-scaling limit therefore lives in the window $1\ll mL\ll(L^2/\GN)^{2/7}$. The mass of the condensate, $ML\simeq N\,mL=\bG(L^2/\GN)(mL)^{-3/2}$, is then parametrically below the black-hole threshold $L^2/\GN$. We also assume that the operator dual to $\Psi$, of dimension $\Delta\simeq mL$, lies below the gap to higher-spin and stringy states, as in any treatment of a heavy bulk field in classical gravity. The numerics use the scaling invariance described below and do not depend on the value of $L^2/\GN$.

\subsection{Dimensionless form and the parameter map}

In trap units (length $\ell$, energy $\omega_t$, $\int|\tilde\varphi|^2=1$), Eq.~\eqref{eq:GPP} is the GPP equation in the convention of becGPP~\cite{becGPP}:
\begin{multline}
\tilde\mu\tilde\varphi=\Big[-\tfrac12\tilde\nabla^2+\tfrac12\tilde r^2+\beta_2|\tilde\varphi|^2+\beta_3|\tilde\varphi|^4\\
-\bG\!\int\!\frac{|\tilde\varphi(\tilde{\bm x}')|^2}{|\tilde{\bm x}-\tilde{\bm x}'|}\,\dd^3\tilde x'\Big]\tilde\varphi .
\label{eq:GPPdl}
\end{multline}
The relativistic couplings map onto it as
\begin{equation}
\begin{gathered}
\beta_2=\frac{\lambda N}{4\sqrt{mL}},\qquad \beta_3=\frac{\gamma N^2}{8L^2},\\
\bG=\Big(\GN m^2-\frac{q^2}{4\pi}\Big)N\sqrt{mL},
\end{gathered}
\label{eq:dict}
\end{equation}
and the observables as
\begin{equation}
\begin{aligned}
(\omega-m)L&=\tilde\mu+O(\epsilon),\\
ML&=N\big[mL+\tilde E(\beta_2,\beta_3,\bG)\big]+O(N\epsilon),
\end{aligned}
\label{eq:dictobs}
\end{equation}
with $\tilde E$ the GPP energy per particle. For $N\to0$, Eq.~\eqref{eq:dictobs} gives $\tilde\mu=\tfrac32$. The exact value is $\Delta-mL=\tfrac32+\tfrac98\epsilon+O(\epsilon^3)$, which illustrates the size of the correction.

The limit is a double-scaling limit, $mL\to\infty$ at fixed $(\beta_2,\beta_3,\bG)$. The relativistic corrections come from five sources: $p^4/m^3$ terms, the spatial metric $\dd r^2/(1+r^2/L^2)$, post-Newtonian terms $O(\Phi_N)$, $(\omega-m)^2/m$ terms, and $\lambda\psi^2/m^2$. Each is $O(\epsilon)$ at fixed $\beta$'s. In particular, the compactness obeys
\begin{equation}
\frac{\GN M}{R}\sim\frac{\GN m^2 N}{m\ell}=\frac{\bG}{mL}\qquad(q=0),
\label{eq:compact}
\end{equation}
so the self-gravity can be $O(1)$ on the scale of the trap while the geometry stays nearly flat. This is the hierarchy that makes a trapped self-gravitating condensate a weak-field object. Under $\Psi\to\Psi\sqrt{\kappa/\kappa'}$, $\lambda\to\lambda\kappa'/\kappa$, $\gamma\to\gamma(\kappa'/\kappa)^2$, $q^2\to q^2\kappa'/\kappa$, the $\beta$'s are invariant and $N\propto1/\kappa'$. A run at $\kappa=1/2$ therefore represents any $\GN/L^2\ll1$, where $N\sim L^2/\GN$ is large and mean-field theory is justified.

The Einstein--massive-scalar case without self-interaction and without gauge field was obtained in Ref.~\cite{BizonEvninFicek2018}, where the limit is the Schr\"odinger--Newton--Hooke equation. That equation has no Newtonian collapse: in a harmonic trap the $-1/r$ attraction is always balanced by the kinetic energy, so the only instability is the relativistic one. Each coupling added here changes the physics, not only the bookkeeping. The quartic coupling $\lambda<0$ brings in the trapped-attractive fold, a Newtonian instability at small compactness, and with it the collapse diagram of Sec.~\ref{sec:fold}, in which one fold line connects the trapped-attractive, Chavanis--Delfini and Kaup limits and the relativistic folds approach the Newtonian ones as $(mL)^{-1}$. The Maxwell field screens self-gravity, Eq.~\eqref{eq:Geff}. Charged and neutral stars with equal $\Geff$ then share one GPP branch; the screening itself is the familiar balance of charged boson stars~\cite{JetzerVanDerBij1989}, and what is new is its place in the dictionary. The sextic coupling $\gamma$ adds a self-bound droplet branch whose liquid--gas end point is moved by gravity (Appendix~\ref{sec:droplet}). The energy map~\eqref{eq:dictobs} turns the GPP energy into the scaling dimension of the charge-$N$ operator of the dual CFT. Finally, the dual state carries the entanglement observables of Sec.~\ref{sec:ee}, which have no counterpart in the derivation of Ref.~\cite{BizonEvninFicek2018} or in the other routes of Table~\ref{tab:routes}. Compared with those routes, the small parameter is geometric, the trap frequency is fixed, and all GPP couplings are fixed by the bulk Lagrangian. We verify the $O(\epsilon)$ convergence on fully back-reacted, nonlinear solutions (Sec.~\ref{sec:results}).

% ======================================================================
\section{Collapse as a fold}
\label{sec:fold}

\subsection{Saddle-node structure}

Fix a ray $(\beta_2,\beta_3,\bG)=(r_b t,s_b t^2,g_r t)$ with $t\propto N$. Along a ray, the ground-state branch of Eq.~\eqref{eq:GPPdl} ends where the $\ell=0$ Bogoliubov operator acquires a zero mode. Write $L_\pm$ for the real and imaginary parts of the linearized GPP operator at a real ground state $\tilde\varphi_0$:
\begin{equation}
\begin{aligned}
L_-&=H_0-\tilde\mu,\\
L_+&=L_-+2\beta_2\tilde\varphi_0^2+4\beta_3\tilde\varphi_0^4+2\tilde\varphi_0\,\mathcal K\,\tilde\varphi_0,
\end{aligned}
\label{eq:Lpm}
\end{equation}
where $\mathcal K$ is the linearized Newtonian kernel. Collective frequencies follow from $\omega^2\in\mathrm{spec}(L_-L_+)$. The phase mode $L_-\tilde\varphi_0=0$ is removed. At a generic fold $t_c$, $L_+$ restricted to $\tilde\varphi_0^\perp$ develops a simple zero eigenvalue. Lyapunov--Schmidt reduction then gives the normal form $t_c-t\propto(\tilde\mu-\tilde\mu_c)^2$, hence
\begin{equation}
\omega_B^2\propto(t_c-t)^{1/2},\qquad \frac{\dd\langle\tilde r\rangle}{\dd t}\propto(t_c-t)^{-1/2}.
\label{eq:foldexp}
\end{equation}
This is the Vakhitov--Kolokolov mechanism~\cite{VakhitovKolokolov1973}. For a trapped attractive gas the exponent $\omega_B\propto(t_c-t)^{1/4}$ was found by Ueda and Leggett~\cite{UedaLeggett1998}. The relativistic counterpart is the turning-point theorem for boson stars~\cite{LeePang1989,GleiserWatkins1989,Kusmartsev1991}: the fundamental radial mode changes stability at the first maximum of $N$. By Eq.~\eqref{eq:firstlaw}, $M$ is extremal there too. Neither statement is new. What the dictionary adds is that the two folds are the same object, displaced by $O(\epsilon)$.

\subsection{The collapse diagram}

Two limits of the GPP fold line $\beta_{2c}(\bG)$ are known in closed numerical form. At $\bG=0$ the trapped attractive gas collapses at $N|a|/a_{\rm ho}=0.575$~\cite{Ruprecht1995}, that is $\beta_{2c}=-4\pi\times0.575$. For $\bG\to\infty$ the trap becomes irrelevant. With $g_2=4\pi a_s/m$, the Chavanis--Delfini maximum mass $M_{\max}=1.012/\sqrt{\GN m|a_s|}$~\cite{ChavanisDelfini2011} becomes
\begin{equation}
\beta_{2c}\,\bG\;\longrightarrow\;-4\pi(1.012)^2\simeq-12.87 .
\label{eq:CD}
\end{equation}
The full line, computed in Sec.~\ref{sec:results}, interpolates between these two limits. Rays with $\beta_2\ge0$ and $\beta_3\ge0$ have no Newtonian fold, because Newtonian gravity with repulsive or vanishing contact interactions cannot overcome the kinetic pressure~\cite{Lieb1977}. Their collapse is relativistic. For $\beta_2=0$ and $q=0$, $\bG=\GN m^2N\sqrt{mL}$, so the Kaup limits $\GN M_{\max}m\to0.633$ and $\GN m^2N_{\max}\to0.653$~\cite{Kaup1968,RuffiniBonazzola1969} become
\begin{equation}
t_{\rm fold}\simeq0.653\,\sqrt{mL}\,\big[1+o(1)\big],
\label{eq:kaup}
\end{equation}
valid once the star is small compared with $\ell$. For $\beta_2>0$ the fold is the Colpi--Shapiro--Wasserman maximum~\cite{Colpi1986}. In particular, the $N_{\max}\propto\sqrt\lambda$ scaling at repulsive coupling belongs to the relativistic fold~\cite{Colpi1986}, not to the Newtonian one~\cite{Chavanis2011}, and at finite $mL$ it is corrected by the trap in the same way as Eq.~\eqref{eq:kaup}.

% ======================================================================
\section{Holographic entanglement}
\label{sec:ee}

Set $L=1$. For a polar cap $A=\{\theta\le\theta_0\}$ on the boundary $S^2$, the RT surface~\cite{RyuTakayanagi2006,HRT2007} lies on the $t=0$ slice, where the metric is $\dd r^2/f+r^2\dd\Omega_2^2$ and $\chi$ drops out. We use $c_0=\cos\theta_0$, $s_0=\sin\theta_0$, and $\delta S=S-S_{\rm vac}$.

\subsection{Exact hemisphere entropy}
\label{sec:hemi}

For $\theta_0=\pi/2$ the equatorial disk is extremal by reflection symmetry, and
\begin{equation}
\delta S_{\rm eq}=\frac{\pi}{2\GN}\int_0^\infty\!\dd r\;r\Big[f^{-1/2}-(1+r^2)^{-1/2}\Big],
\label{eq:Seq}
\end{equation}
which is UV finite because $f-f_0=O(r^{-1})$. The disk is the RT surface only if it is minimal. Positivity of the boundary relative entropy, $\Delta H-\delta S\ge0$~\cite{Blanco2013,Lashkari2016}, therefore gives a sufficient test: wherever $\delta S_{\rm eq}>\Delta H$, the true RT surface must break the reflection symmetry. A sharper, local test is the stability of the disk as a minimal surface. The disk is totally geodesic, so its Jacobi operator is $J=-\Delta_\Sigma-\mathrm{Ric}(n,n)$, with $\mathrm{Ric}(n,n)=-\tfrac32-(f-1)/2r^2+\tfrac{\kappa}{2}\rhoE$ for the spatial metric $\dd r^2/f+r^2\dd\Omega^2$. A negative eigenvalue in the reflection-odd, axisymmetric sector signals a pitchfork of extremal surfaces. Beyond it the equatorial disk no longer computes the entropy. Multiple extremal surfaces and entanglement shadows in compact-star geometries are known~\cite{Hubeny2012,Nogueira2013,Freivogel2015}; what we compute is where the Jacobi threshold lies relative to the turning point of the star.

\subsection{Modular energy minus relative entropy}

At linear order in $\GN$ the entropy of any cap obeys
\begin{equation}
\delta S(\theta_0)=\Delta H(\theta_0)-K(\theta_0),
\label{eq:split}
\end{equation}
which is the holographic first law with a bulk source~\cite{Blanco2013,Faulkner2014,JLMS2016}. Here the modular energy is
\begin{equation}
\Delta H=2\pi\!\int_A\!\dd\Omega\,\frac{\cos\theta-c_0}{s_0}\,\langle T_{tt}\rangle=\frac{\pi M(1-c_0)^2}{2s_0},
\label{eq:dH}
\end{equation}
since $\langle T_{tt}\rangle=M/4\pi$. The second term,
\begin{equation}
K=\int_{\Sigma_A}\!\dd^3x\;\xi^t\rhoE,\qquad \xi^t=\frac{2\pi}{s_0}\,\frac{r\cos\theta-c_0\sqrt{1+r^2}}{\sqrt{1+r^2}},
\label{eq:K}
\end{equation}
is the Killing energy of the bulk matter in the entanglement wedge $\Sigma_A$ ($\xi^t>0$). More precisely, $S_A=A[\gamma_A]/4\GN+S_{\rm bulk}$~\cite{FaulknerLewkowyczMaldacena2013}, boundary relative entropy equals bulk relative entropy~\cite{JLMS2016}, and the bulk modular Hamiltonian of the wedge is the boost charge~\cite{CasiniHuertaMyers2011}, so that $\Delta H-\delta S=K-\Delta S_{\rm bulk}$. For a coherent condensate the $O(N^0)$ term $\Delta S_{\rm bulk}$ vanishes, and $K$ is the boundary relative entropy $\Srel(\rho_A\|\rho_A^{\rm vac})$. For spherical $\rhoE$ the angular integral is elementary:
\begin{equation}
K=\frac{4\pi^2}{s_0}\!\int_0^\infty\!\dd r\,\frac{r^3\rhoE}{\sqrt{1+r^2}}\,W(a),\qquad a=\frac{c_0\sqrt{1+r^2}}{r},
\label{eq:Kradial}
\end{equation}
with $W=\tfrac12(1-a)^2$ for $|a|\le1$, $W=-2a$ for $a<-1$, and $W=0$ for $a>1$. For the hemisphere, Eq.~\eqref{eq:split} follows in one line. The linearization of Eq.~\eqref{eq:Seq} is $\delta S_{\rm lin}=\tfrac{\pi}{2}\int m(r)(1+r^2)^{-3/2}\dd r$, and integrating by parts with $\int_0^r(1+s^2)^{-3/2}\dd s=r/\sqrt{1+r^2}$ gives $\tfrac{\pi}{2}M-2\pi^2\int r^3\rhoE(1+r^2)^{-1/2}\dd r=\Delta H-K$. Two exact identities follow from Eqs.~\eqref{eq:dH}--\eqref{eq:Kradial}. One is purity, $\delta S(\theta_0)=\delta S(\pi-\theta_0)$, which holds because $W(-a)-W(a)=2a$. The other is $\delta S_{\rm lin}=\Delta H-K$. Both serve as numerical gates (Sec.~\ref{sec:num}).

\subsection{Cap entropies as Radon data}
\label{sec:radon}

Equation~\eqref{eq:split} holds for a cap $A(\bom,\theta_0)=\{\bm n:\bom\cdot\bm n>c_0\}$ of any orientation, for non-spherical matter, and for matter that is not static. At linear order in $\GN$ the RT surface of every cap is the vacuum one, since a first-order displacement of an extremal surface leaves its area unchanged; the entropy changes only through the metric perturbation integrated over the vacuum surface. For any linearized solution sourced by $T_{\mu\nu}$, the Iyer--Wald identity for the Killing vector $\xi$ that generates the modular flow of the wedge~\cite{IyerWald1994,Faulkner2014,SwingleVanRaamsdonk2014}, integrated over the part $\Sigma_A$ of a global time slice between the cap and its RT surface, gives $\delta S=\Delta H-\int_{\Sigma_A}\xi^\mu T_{\mu\nu}n^\nu\dd\Sigma$. On that slice $\xi$ is normal to $\Sigma_A$ (Appendix~\ref{app:klein}), so momentum densities and stresses drop out, and $K=\int_{\Sigma_A}\xi^t\rhoE\,\dd^3x$ with the flat coordinate measure $\dd^3x=r^2\dd r\,\dd\Omega$. Here $\rhoE$ is the energy density of all bulk fields measured by static observers, the electrostatic energy included. Introduce Klein (Beltrami) coordinates
\begin{equation}
\by=\frac{\bm x}{\sqrt{1+r^2}},\qquad |\by|<1 .
\label{eq:klein}
\end{equation}
The weight of Eq.~\eqref{eq:K}, rotated to the axis $\bom$, becomes affine,
\begin{equation}
\xi^t=\frac{2\pi}{s_0}\,\big(\bom\cdot\by-c_0\big),
\label{eq:xiklein}
\end{equation}
and every RT surface of a cap is the plane $\bom\cdot\by=c_0$. With $\dd^3x=(1-|\by|^2)^{-5/2}\dd^3y$ and the boundary modular weight $2\pi(\bom\cdot\bm n-c_0)/s_0$~\cite{CasiniHuertaMyers2011}, Eq.~\eqref{eq:split} reads
\begin{multline}
\frac{s_0}{2\pi}\,\delta S(\bom,c_0)=\int_{S^2}\!\dd\Omega\,(\bom\cdot\bm n-c_0)_+\langle T_{tt}(\bm n)\rangle\\
-\int_{|\by|<1}\!\dd^3y\,(\bom\cdot\by-c_0)_+\,\nu(\by),
\label{eq:radon}
\end{multline}
with $\nu=\rhoE(1-|\by|^2)^{-5/2}$. Both terms carry the same kernel. The boundary energy density is a shell at $|\by|=1$, and the bulk energy density fills the ball. Since $\partial_{c}^2(\bom\cdot\by-c)_+=\delta(\bom\cdot\by-c)$, $\partial_{c_0}^2[s_0\,\delta S]/2\pi$ is the Euclidean Radon transform of $\langle T_{tt}\rangle\,\delta(|\by|-1)-\nu$.

This is the global-$\AdS_4$ form of the statement of Ref.~\cite{Lin2015}, that a second derivative of the relative entropy is the integral of the energy density over the RT surface. That the totally geodesic Radon transform on hyperbolic space becomes a weighted Euclidean one in the Klein model is known in integral geometry~\cite{Kurusa1991,BerensteinTarabusiKurusa1997}. The Radon transform is injective on compactly supported distributions, and in three dimensions its inverse is local, $f(\by)=-(8\pi^2)^{-1}\oint\dd\bom\,(\partial_c^2\mathcal Rf)(\bom,\bom\cdot\by)$~\cite{Helgason1999,Natterer2001}. The second $c$ derivative of the data~\eqref{eq:radon} is the transform of $\langle T_{tt}\rangle\,\delta(|\by|-1)-\nu$, a distribution supported in the closed ball. Its exact inverse, evaluated at $|\by|<1$, is therefore $-\nu(\by)$: the shell drops out of the interior identically, whatever $\langle T_{tt}\rangle$ is. Appendix~\ref{app:klein} checks this harmonic by harmonic. Hence
\begin{multline}
\rhoE(\by)=\frac{(1-|\by|^2)^{5/2}}{16\pi^3}\\
\times\oint_{S^2}\!\dd\bom\;\frac{\partial^4}{\partial c^4}\Big[\sqrt{1-c^2}\;\delta S(\bom,c)\Big]_{c=\bom\cdot\by}.
\label{eq:inversion}
\end{multline}
The relative entropy of Ref.~\cite{Lin2015} is $\Delta H-\delta S$, and its inversion needs $\langle T_{tt}\rangle$ as a separate input. Equation~\eqref{eq:inversion} acts on the entanglement entropies themselves, and the operator $\oint\dd\bom\,\partial_c^4$ annihilates the modular-energy term at every interior point. The idea that a fixed operator on the entropies removes the boundary stress tensor and leaves the bulk sources is that of kinematic space, where the wave operator on the space of caps plays this role~\cite{deBoerHellerMyersNeiman2016,CzechLamprouMcCandlishSully2015,CzechEtAl2017}. Equation~\eqref{eq:inversion} is an explicit realization for caps of $S^2$ in global $\AdS_4$: the Klein coordinates reduce the problem to the Euclidean Radon transform, the operator and the inversion are in closed form, and the hypotheses below are stated so that the formula can be tested and its error measured. The stress tensor is of course also encoded in the entropies: small caps obey the first law $\delta S\to\Delta H$, and the distributional inverse returns $\langle T_{tt}\rangle$ as the shell on $|\by|=1$. What Eq.~\eqref{eq:inversion} adds is that no separation of the two terms is needed, neither a subtraction nor an extrapolation to small caps. The drop-out uses only the support of the shell, not the dimension of the slice. Three dimensions add locality: $\nu(\by)$ needs only the entropies of caps whose RT planes pass through $\by$ or near it.

The hypotheses of Eq.~\eqref{eq:inversion} are four. (i)~Linear order in $\GN$ about global $\AdS_4$, so that every RT surface is a vacuum plane. (ii)~Classical bulk matter, with the $O(N^0)$ bulk entanglement $\Delta S_{\rm bulk}$ neglected. (iii)~Entropies of caps of all orientations $\bom\in S^2$ and all $|c|<1$, anchored at one boundary time. (iv)~Integrability of $\nu$ on the Klein ball, $\int\nu\,\dd^3y=\int\rhoE\,\dd^3x<\infty$, which holds for any finite-energy configuration with $\Delta>3/2$; the Coulomb tail of a charged condensate gives $\nu\propto(1-|\by|)^{-1/2}$, still integrable. No symmetry, no staticity, and no smallness of the matter compared with $L$ is assumed. Under (i)--(iv) the cap entropies determine $\rhoE$ uniquely, at every depth. The error of (i) is measured in Sec.~\ref{sec:ctres}. Beyond the Jacobi instability of Sec.~\ref{sec:hemi} the expansion in $\GN$ follows the reflection-symmetric extremal surface, which is then no longer minimal, and (i) fails for near-equatorial caps (Sec.~\ref{sec:rtbound}). Smoothing $\delta S$ over $\Delta c$ blurs $\nu$ isotropically by $\Delta c$ in the Klein ball. Near the center $\by\simeq\bm x$, so a feature of size $\delta$ needs cap angles resolved to $\Delta\theta_0\simeq\delta$.

\subsection{Entanglement tomography in the GPP regime}

In the GPP regime the matter sits at $r=O(\ell)\ll1$. For caps with $z_*\equiv\cot\theta_0=O(\ell)$, Eq.~\eqref{eq:K} expands to $\xi^t=2\pi(z-z_*)[1+O(\epsilon)]$ and $\rhoE=m|\varphi|^2[1+O(\epsilon)]$. In trap units $\tilde z_*=\sqrt{mL}\cot\theta_0$, and Eq.~\eqref{eq:K} becomes
\begin{equation}
\begin{gathered}
\Srel(\theta_0)=2\pi N\sqrt{mL}\,F(\tilde z_*)\,[1+O(\epsilon)],\\
F(\tilde z)=\int(\tilde z'-\tilde z)_+\,\tilde n(\tilde{\bm x}')\,\dd^3\tilde x',
\end{gathered}
\label{eq:Fplanar}
\end{equation}
which is a planar Radon transform of the condensate density $\tilde n=|\tilde\varphi|^2$. Two derivatives remove the kernel, $F''=\ncol$, where $\ncol(\tilde z)=\int\tilde n\,\dd\tilde x\,\dd\tilde y$ is the column density. The modular energy is $O(M)=O(NmL)$, but it depends on the cap only through $c_0$, and $\partial_{\tilde z_*}=O((mL)^{-1/2})\,\partial_{c_0}$. Hence $\partial^2_{\tilde z_*}\Delta H=O(N)$ for any $\langle T_{tt}\rangle$ of order $M/4\pi$, spherical or not, while $\Srel=O(N\sqrt{mL})$:
\begin{equation}
\frac{\partial^2\delta S}{\partial\tilde z_*^2}=-2\pi N\sqrt{mL}\;\ncol(\tilde z_*)\;+\;O(N).
\label{eq:tomo}
\end{equation}
The boundary entanglement entropy of a family of parallel caps thus reconstructs the column density of the bulk condensate along $\bom$. This is the quantity an absorption image records in a cold-atom experiment~\cite{Ketterle1999}. The set of such images over all $\bom$ is inverted by Eq.~\eqref{eq:inversion}. At $r=O(\ell)$ one has $\by=\bm x[1+O(\epsilon)]$ and $c=\ell\tilde z_*[1+O(\epsilon)]$, so $\partial_c^4=(mL)^2\partial^4_{\tilde z_*}$, and
\begin{multline}
|\tilde\varphi(\tilde{\bm x})|^2=\frac{1}{16\pi^3N\sqrt{mL}}\oint\dd\bom\;\partial^4_{\tilde z_*}\,\delta S(\bom,\tilde z_*)\Big|_{\tilde z_*=\bom\cdot\tilde{\bm x}}\\
\times\big[1+O(\epsilon)\big].
\label{eq:gppct}
\end{multline}
In Eq.~\eqref{eq:gppct} the shell no longer cancels exactly, because the factor $\sqrt{1-c^2}$ has been set to one. For a modular energy of order $M$ the residue in $\partial^4_{\tilde z_*}\delta S$ is $O(N\epsilon)$, against $O(N\sqrt{mL})$ for the condensate term, a relative $O(\epsilon^{3/2})$ inside the stated error. The planar limit~\eqref{eq:gppct} is reliable only when the condensate is small compared with $L$. Section~\ref{sec:ctres} shows that it fails once $\langle r^2\rangle^{1/2}\sim L$, while Eq.~\eqref{eq:inversion} does not. For the hemisphere and a spherical condensate, $\Srel=\pi mN\langle|z|\rangle=\tfrac{\pi}{2}mN\langle r\rangle$, so
\begin{equation}
\delta S_{\rm hemi}=\frac{\pi}{2}\big(ML-mN\langle r\rangle\big)\,[1+O(\GN,\epsilon)],
\label{eq:hemiNR}
\end{equation}
and the size of the condensate is an entanglement observable.

\subsection{Entanglement susceptibility at the fold}

Along a ray $N\propto t$, and Eq.~\eqref{eq:hemiNR} gives
\begin{equation}
\frac{\partial\,\delta S_{\rm hemi}}{\partial N}=\frac{\pi}{2}\Big[\omega L-\sqrt{mL}\,\chi_E\Big],\quad \chi_E\equiv\frac{\dd(t\langle\tilde r\rangle)}{\dd t}.
\label{eq:chiE}
\end{equation}
The modular part is analytic at the fold, since $\dd M/\dd N=\omega$ by Eq.~\eqref{eq:firstlaw}. The relative-entropy part is not. Equation~\eqref{eq:foldexp} gives $\chi_E\propto(t_c-t)^{-1/2}$, so
\begin{equation}
\chi_E\,\tilde\omega_B^2\;\longrightarrow\;\mathcal A(r_b,s_b,g_r)\quad(t\to t_c^-),
\label{eq:A}
\end{equation}
a finite, ray-dependent amplitude. The exponent itself is generic: at a saddle-node every observable that depends smoothly on the state has a derivative with respect to $N$ that diverges as $(N_c-N)^{-1/2}$, and $\chi_E$ is, by Eq.~\eqref{eq:hemiNR}, the susceptibility of the mean radius. What the holographic description adds is the split between boundary observables. The modular energy is fixed by the conserved charges, and its first derivative stays finite at the fold, $\dd M/\dd N=\omega$. The divergent first derivative is carried by non-conserved observables, among them the condensate $\langle\mathcal O\rangle$ and the cap entropies. Among these, the entropies are the ones whose singular part is geometric: it is the bulk relative entropy, a radial moment of the bulk energy density, which becomes infinitely sensitive to $N$ exactly where the breathing mode softens. Entanglement observables themselves are smooth functions of $\psi_0$ across the fold; only their derivatives with respect to $N$ diverge.

% ======================================================================
\section{Numerical method}
\label{sec:num}

All computations run on a GPU in double precision.

\paragraph*{GPP.}
Ground states are spherically symmetric. We write $\tilde\varphi=u(\tilde r)/\tilde r$ and expand $u$ in a sine series on $[0,R]$ (DST-I, $J=320$). The kinetic term is then diagonal and spectrally accurate. The Newtonian potential is computed by solving $w''=4\pi\bG u^2/\tilde r$ for $w=\tilde r U$ in the same basis, with $w(R)=-\bG$. With this choice the discrete virial identity $2E_{\rm kin}-2E_{\rm trap}+E_G+3E_2+6E_3=0$ holds to $10^{-11}$, against $10^{-4}$ for a direct radial quadrature of the Green function. Stationary states are found by Newton's method on $(u,\tilde\mu)$ with the norm constraint bordered into the Jacobian, and are continued along rays with adaptive steps and automatic regridding. The $\ell=0$ spectrum is obtained from the symmetric matrix $L_-^{1/2}L_+L_-^{1/2}$, with the phase mode deflated. The fold is located by fitting $t=t_c-c_1\tilde\omega_B^4-c_2\tilde\omega_B^8$ as $\tilde\omega_B\to0$. Imaginary-time gradient flow with discrete normalization (GFDN) and Strang splitting~\cite{BaoDu2004,BaoCai2013} supplies initial data and serves as an independent check of the GPU path.

\paragraph*{Fold amplitude.}
The amplitude~\eqref{eq:A} is evaluated at the fold itself. The chemical potential stays a regular parameter through the saddle-node, so we continue the branch in $\mu$. At fixed $\mu$ the unknowns are $(u,t)$, and the bordered Jacobian $\big[\begin{smallmatrix}L_+ & (\partial_t\mathcal H)u\\ 2(\mathsf Wu)^{\mathsf T} & 0\end{smallmatrix}\big]$ stays regular at the fold; here $\mathcal H$ is the GPP operator and $\mathsf W$ the diagonal of quadrature weights. Then $t(\mu)$, $X=t\langle\tilde r\rangle$ and $\tilde\omega_B^2$ are smooth, and at $t'(\mu_c)=0$
\begin{equation}
\mathcal A=\frac{X'(\mu_c)\,(\tilde\omega_B^2)'(\mu_c)}{t''(\mu_c)} .
\end{equation}
We fit Chebyshev polynomials of degree 6 and 8 on 13 nodes, on windows with $(t_c-t)/t_c\le10^{-3}$, $3\times10^{-4}$ and $10^{-4}$, and repeat with $J=480$. As a check of the construction, $\tilde\omega_B^2$ vanishes at the computed $\mu_c$ to $2\times10^{-8}$ of its range on every ray.

\paragraph*{AdS stars.}
Equations~\eqref{eq:KG}--\eqref{eq:f} are integrated as eleven first-order ODEs. Besides the six fields, the state carries $N_r$, $m(r)$ and the integrands of Eqs.~\eqref{eq:Seq}, \eqref{eq:Kradial} and $\delta S_{\rm lin}$. We use RK4 in $x=r\sqrt{mL}$ (step $0.004$, $x\le12$) from a series start at the origin. The frequency is found by multisection on $\Omega(0)$ in the frame $\chi(0)=0$, with $48$ candidates per round and $7$ rounds, giving a bracket width of $10^{-11}$. A candidate is classified as above the ground state if $\psi$ crosses zero and as below it if $\psi$ turns upward. By Sturm comparison this classification is monotonic. All amplitudes and configurations are shot in a single batch. A final pass freezes the state at the turning point $r_*$, and the exterior is added analytically from Eq.~\eqref{eq:MN}. A second pass refines the amplitude grid around the first maximum of $N$.

\paragraph*{Three-dimensional states and tomography.}
Rotating states solve Eq.~\eqref{eq:GPPdl} with $\beta_3=0$ and the term $-\Omega_{\rm rot} L_z$ on a $128^3$ periodic grid on $[-8,8)^3$ ($\Delta x=0.125$), by imaginary-time Strang splitting with normalization and time steps $10^{-2}$, $4\times10^{-3}$ and $1.5\times10^{-3}$~\cite{BaoCai2013}. The production grid is converged to a few parts in $10^5$ in the energy, angular momentum and radius, and to $0.03$ lattice spacings in the vortex positions (Appendix~\ref{app:conv}). The rotation term is split along alternating directions, $\exp[-\tau(k_x^2/2+\Omega_{\rm rot} yk_x)]$ with an FFT in $x$ and $\exp[-\tau(k_y^2/2-\Omega_{\rm rot} xk_y)]$ with an FFT in $y$; the $k_z^2/2$ term is applied with an FFT in $z$, and the potential in two half steps in position space. The Newtonian potential uses the zero-padded Hockney convolution with the cell-averaged kernel $2.380077/\Delta x$ at the origin~\cite{HockneyEastwood1988}. Vortex lattices are seeded on a triangular lattice of areal density $\Omega_{\rm rot}/\pi$ and relaxed for up to $2000$ steps at each time step, stopping early when the energy changes by less than a relative $10^{-11}$ over $500$ steps. Imaginary time lowers the rotating-frame energy $E-\Omega_{\rm rot}L_z$ monotonically, so the result is a stationary state and a local minimum of that energy; we did not search for competing lattice arrangements. The forward model evaluates Eq.~\eqref{eq:radon} exactly at leading order. For each of $N_\omega$ Fibonacci orientations, the projections $\bom\cdot\by_i$ of all grid points are binned on a grid $c_j$ of spacing $\ell\Delta x/4$ with cloud-in-cell weights. The histogram is smoothed by a Gaussian of width $\sigma=0.7\,\ell\Delta x$, which defines the resolution kernel, and the kernel $(\bom\cdot\by-c)_+$ is applied by two reversed cumulative sums. This is exact for piecewise-linear histograms, and it preserves the purity identity $Q(\bom,c)-Q(-\bom,-c)=\langle\bom\cdot\by\rangle-c$, with $Q(\bom,c)=\int(\bom\cdot\by-c)_+\nu\,\dd^3y$ for unit-normalized $\nu$, to machine precision. A boundary stress tensor with $\ell=0,2,4,6$ components is added to the data. It is chosen freely, not computed from the state, so that the drop-out of the shell is tested on a non-spherical $\langle T_{tt}\rangle$. The inversion applies a five-point $\partial_c^4$, back-projects with linear interpolation in $c$ and equal Fibonacci weights, and maps $\nu$ back to $\tilde n$. The planar comparison resamples the same data onto $\tilde z_*$ with quintic splines and applies Eq.~\eqref{eq:gppct}.

\paragraph*{Error measures.}
Since the exact reconstruction from smoothed data is $\nu$ convolved with the Gaussian of width $\sigma$, the reference density $\tilde n_{\rm ref}$ is the true $\tilde n$ convolved with an isotropic Gaussian of width $\sigma/\ell=0.7\,\Delta x=0.0875$ in trap units. The kernel is isotropic in $\by$, the reference in $\tilde{\bm x}$; the two agree up to $O(\langle r^2\rangle/L^2)$. The cloud is the set $\mathcal C=\{\tilde n_{\rm ref}>10^{-3}\max\tilde n_{\rm ref}\}$. We quote three numbers, all evaluated on the full $128^3$ grid: the on-cloud error $\|\tilde n_{\rm rec}-\tilde n_{\rm ref}\|_{\mathcal C}/\|\tilde n_{\rm ref}\|_{\mathcal C}$, the leakage $\|\tilde n_{\rm rec}-\tilde n_{\rm ref}\|_{\overline{\mathcal C}}/\|\tilde n_{\rm ref}\|$, and the global error $\|\tilde n_{\rm rec}-\tilde n_{\rm ref}\|/\|\tilde n_{\rm ref}\|$, with $\|\cdot\|$ the discrete $L^2$ norm. Vortices are located in the plane $\tilde z=0$. A grid point is a vortex if it is the minimum of its $5\times5$ neighborhood, if the mean density over its $9\times9$ neighborhood exceeds one quarter of the maximum in the plane, and if its own density is below $0.6$ of that mean. The reference vortices are those of $\tilde n_{\rm ref}$. A reference vortex is recovered if a vortex of the reconstruction lies within half a lattice spacing, $a/2$ with $a=(2\pi/\sqrt3\,\Omega_{\rm rot})^{1/2}=2.13$, and the position error is the rms distance of matched pairs in units of $a$. One grid cell is $\Delta x=0.059\,a$.

For the back-reacted stars, $r_{99}$ is the radius that encloses $99\%$ of the charge $N$. The entropy error $\epsilon_S$ is $\max_c|\delta S(c)-\delta S_{\rm lin}(c)|/K_{\rm hemi}$ over caps with $|c|\le c_{99}=r_{99}/\sqrt{1+r_{99}^2}$, where $\delta S$ is the exact RT entropy, $\delta S_{\rm lin}=\Delta H-K$ the leading-order one evaluated on the true $\rhoE$, and $K_{\rm hemi}$ the relative entropy of the hemisphere. The image error $\epsilon_{\rm img}$ is the radial $L^2$ norm, with weight $r^2$ on $0.08\,r_{99}\le r\le r_{99}$, of the difference between Eq.~\eqref{eq:inversion} applied to exact and to leading-order entropies, relative to the latter. It isolates the truncation of Eq.~\eqref{eq:radon}; the same inversion applied to leading-order entropies reproduces the true $\rhoE$ to $0.2$--$2.3\%$.

\paragraph*{Gates.}
Before any production run, the code checks 30 identities and benchmarks (Table~\ref{tab:gates}), and 26 further checks accompany the fold amplitudes, the exact RT surfaces and the grid convergence. Among them are the first law~\eqref{eq:firstlaw}, equality of the Einstein and Noether masses, the vacuum limit $\omega\to\Delta/L$, the identity $\delta S_{\rm lin}=\Delta H-K$, purity, the Ruprecht threshold, and the Schr\"odinger--Newton constants $\tilde E=-0.0543\,\bG^2$ and $\tilde\mu=-0.163\,\bG^2$~\cite{Membrado1989}, and, for the tomography, the forward model against direct summation of $\xi^t\rhoE$ in global coordinates and the cancellation of the boundary stress tensor.

\begin{table*}[t]
\caption{Validation gates.}
\label{tab:gates}
\begin{ruledtabular}
\begin{tabular}{lcc}
check & reference & rel.\ error \\ \hline
HO $\tilde\mu$, $\tilde\omega_B$ & $3/2$, $2$ & $3\times10^{-15}$, $1\times10^{-10}$ \\
TF breathing ($\beta_2=4000$) & $\sqrt5$ & $2.3\times10^{-3}$\,\footnote{Finite-$N$ correction to the Thomas--Fermi value.} \\
GPU imaginary time vs Newton, $\tilde E$ & --- & $\le8.2\times10^{-8}$ \\
$N|a|/a_{\rm ho}$ at collapse & $0.575$~\cite{Ruprecht1995} & $5.9\times10^{-5}$ ($0.57497$) \\
fold exponent & $1/4$ & $3.7\times10^{-6}$ \\
Schr\"odinger--Newton $\tilde E/\bG^2$ & $-0.0543$~\cite{Membrado1989} & $8.0\times10^{-4}$ ($-0.054256$) \\
$\omega(\psi_0\to0)$ & $\Delta$, Eq.~\eqref{eq:Delta} & $\le2.0\times10^{-10}$ \\
$\dd M-\omega\,\dd N$ & $0$ & $\le2.2\times10^{-10}$ \\
$M_{\rm Einstein}-M_{\rm Noether}$ & $0$ & $\le2.4\times10^{-12}$ \\
$\delta S_{\rm lin}-(\Delta H-K)$ & $0$ & $\le1.5\times10^{-10}$ \\
$\delta S(\theta_0)-\delta S(\pi-\theta_0)$ & $0$ & $\le3.8\times10^{-7}$ \\
Hockney potential, Gaussian & $-\mathrm{erf}(r/\sqrt2)/r$ & $3.4\times10^{-4}$ \\
free gas, $\Omega_{\rm rot}=0.5$: $E$, $\langle L_z\rangle$ & $3/2$, $0$ & $7\times10^{-12}$, $2\times10^{-12}$ \\
3D vs radial, $\Omega_{\rm rot}=0$ & --- & $1.3\times10^{-3}$\,\footnote{Second-order Hockney kernel, $72^3$ grid.} \\
Klein forward model, $K$ & direct sum & $\le8\times10^{-12}$ \\
forward-model purity & $0$ & $4\times10^{-14}$ \\
round trip, Eq.~\eqref{eq:inversion} & Gaussian & $3.4\times10^{-3}$ \\
$\langle T_{tt}\rangle$ drops out ($\ell\le6$) & $0$ & $5.5\times10^{-5}$ \\
exact RT, vacuum caps & $\delta S=0$ & $\le6\times10^{-12}$ \\
exact RT vs Eq.~\eqref{eq:Seq}, hemisphere & --- & $4.6\times10^{-8}$ \\
$\tilde\omega_B^2$ at $t'(\mu_c)=0$ & $0$ & $\le1.7\times10^{-8}$ \\
\end{tabular}
\end{ruledtabular}
\end{table*}

% ======================================================================
\section{Results}
\label{sec:results}

\subsection{Collapse diagram}

\begin{figure}[t]
\includegraphics[width=\columnwidth]{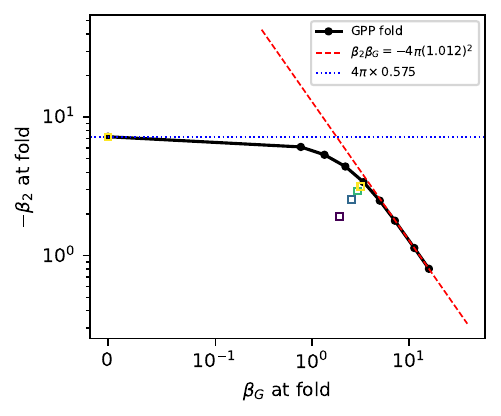}
\caption{GPP fold line (dots) between the trapped-attractive limit $-\beta_{2c}=4\pi\times0.575$ (dotted) and the Chavanis--Delfini hyperbola~\eqref{eq:CD} (dashed). Open squares: folds of the back-reacted AdS stars on the same rays, $mL=16,\dots,128$.}
\label{fig:collapse}
\end{figure}

\begin{table}[t]
\caption{Folds on GPP rays $(\beta_2,\bG)=(r_bt,t)$ and on the contact ray ($\bG\equiv0$). The amplitude $\mathcal A$ of Eq.~\eqref{eq:A} is evaluated at the fold by continuation in $\mu$ (Sec.~\ref{sec:num}); it is stable to $2\times10^{-7}$ under changes of the fit window and the radial grid.}
\label{tab:folds}
\begin{ruledtabular}
\begin{tabular}{ccccc}
$r_b$ & $\bG{}_c$ & $\beta_{2c}$ & $\omega_B$ exp. & $\mathcal A$ \\ \hline
contact & 0 & $-7.2252$ & $0.25000$ & $-3.270$ \\
$-8$ & $0.7621$ & $-6.0964$ & $0.24998$ & $-4.272$ \\
$-4$ & $1.3377$ & $-5.3509$ & $0.25003$ & $-5.288$ \\
$-2$ & $2.2006$ & $-4.4012$ & $0.25003$ & $-7.446$ \\
$-1$ & $3.3934$ & $-3.3934$ & $0.24997$ & $-12.50$ \\
$-0.5$ & $4.9862$ & $-2.4931$ & $0.24998$ & $-25.95$ \\
$-0.25$ & $7.1434$ & $-1.7859$ & $0.25006$ & $-64.39$ \\
$-0.1$ & $11.341$ & $-1.1341$ & $0.25008$ & $-243.1$ \\
$-0.05$ & $16.048$ & $-0.8024$ & $0.25009$ & $-682.8$ \\
\end{tabular}
\end{ruledtabular}
\end{table}

Figure~\ref{fig:collapse} and Table~\ref{tab:folds} give the fold line. At $\bG=0$ the fold gives $N|a|/a_{\rm ho}=0.57497$, the trapped-attractive value~\cite{Ruprecht1995}. Along the rays the product $\beta_{2c}\bG{}_c$ is $-4.65$, $-7.16$, $-9.69$, $-11.52$, $-12.43$, $-12.76$ and $-12.86$, and reaches $-12.877$ on the most gravity-dominated ray. This is the Chavanis--Delfini limit~\eqref{eq:CD}, $-12.870$, to $6\times10^{-4}$. All fitted exponents agree with $1/4$ to $10^{-4}$. The amplitude $|\mathcal A|$ grows monotonically, from $3.270$ on the contact ray to $682.8$ at $r_b=-0.05$. Toward the Chavanis--Delfini limit the Schr\"odinger--Newton scaling (lengths $\propto t^{-1}$, frequencies $\propto t^2$, $t_c\propto|r_b|^{-1/2}$) predicts $\mathcal A\propto|r_b|^{-3/2}$; the computed $\mathcal A|r_b|^{3/2}$ is $-8.05$, $-7.69$ and $-7.63$ at $r_b=-0.25$, $-0.1$ and $-0.05$. The relative-entropy singularity sharpens as self-gravity takes over the collapse.

\subsection{Dictionary at $O(1/mL)$}

\begin{figure*}[t]
\includegraphics[width=0.85\textwidth]{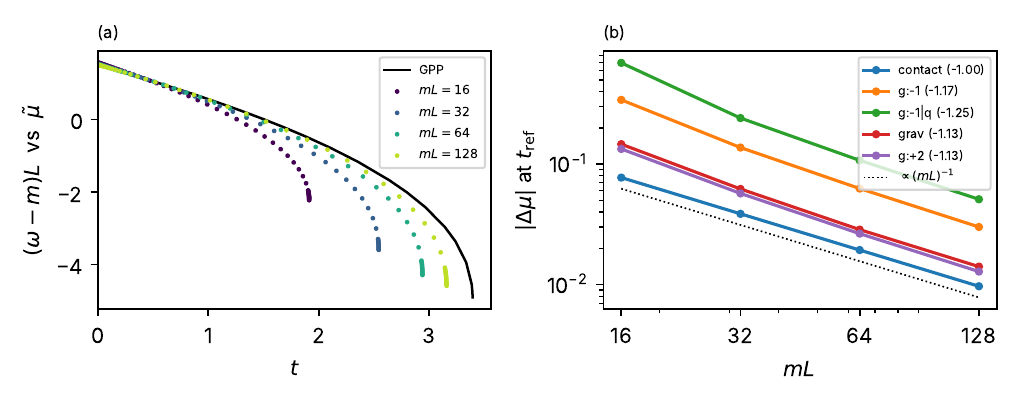}
\caption{(a) $(\omega-m)L$ of back-reacted AdS stars against the GPP chemical potential $\tilde\mu$ on the ray $(\beta_2,\bG)=(-t,t)$. (b) Deviation $|\Delta\tilde\mu|$ at $t_{\rm ref}$ against $mL$ for all rays; dotted line, $(mL)^{-1}$.}
\label{fig:dict}
\end{figure*}

\begin{table}[t]
\caption{Relativistic minus GPP values at $t_{\rm ref}=0.4\,t_c$ on the ray $(-t,t)$, and the relativistic fold.}
\label{tab:dict}
\begin{ruledtabular}
\begin{tabular}{ccccc}
$mL$ & $\Delta\tilde\mu$ & $\Delta\tilde E$ & $\Delta\langle\tilde r\rangle/\langle\tilde r\rangle$ & $t_{\rm fold}$ \\ \hline
16 & $-0.342$ & $-0.0805$ & $-0.135$ & $1.914$ \\
32 & $-0.137$ & $-0.0318$ & $-0.0580$ & $2.544$ \\
64 & $-0.0627$ & $-0.0142$ & $-0.0273$ & $2.940$ \\
128 & $-0.0301$ & $-0.00674$ & $-0.0132$ & $3.158$ \\
GPP & 0 & 0 & 0 & $3.393$ \\
\end{tabular}
\end{ruledtabular}
\end{table}

\begin{figure*}[t]
\includegraphics[width=0.75\textwidth]{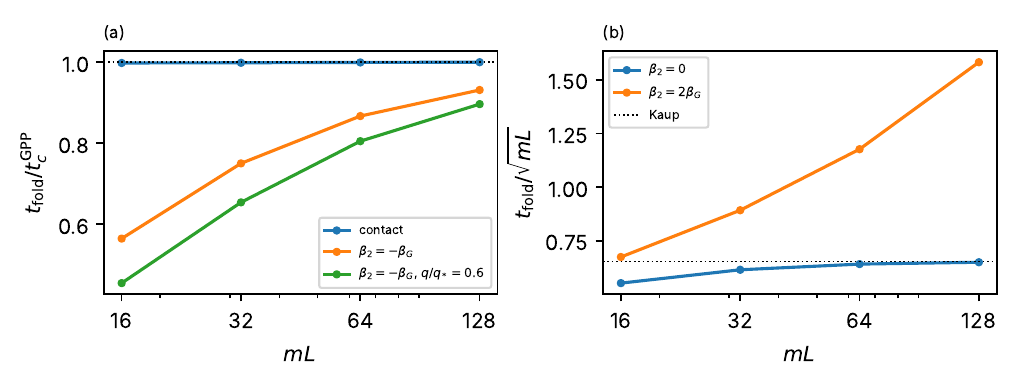}
\caption{Relativistic folds (first maximum of $N$) of back-reacted AdS stars. (a) Rays with a Newtonian fold, normalized to the GPP fold $t_c^{\rm GPP}$. (b) Rays without one, $t_{\rm fold}/\sqrt{mL}$; dotted, the Kaup value $0.653$ of Eq.~\eqref{eq:kaup}.}
\label{fig:relfold}
\end{figure*}

The deviations in Table~\ref{tab:dict} decay as $(mL)^{-p}$. A fit over $16\le mL\le128$ gives $p=0.999$ on the contact ray, $p=1.17$ on the attractive gravitating ray, $p=1.25$ for the charged sequence ($q/q_*=0.6$), and $p=1.13$ on the rays $\beta_2=0$ and $\beta_2=2t$. Between $mL=64$ and $128$ the local exponents are $1.00$, $1.06$, $1.07$, $1.02$ and $1.05$, in the same order, so the asymptotic rate is the predicted $O(1/mL)$. The charged and neutral sequences with equal $\Geff$ converge to the same GPP branch: their deviations at $mL=128$ are $-0.051$ and $-0.030$, both decreasing as $1/mL$. This confirms the screening~\eqref{eq:Geff}. On every ray the relativistic fold lies below the Newtonian one [Fig.~\ref{fig:relfold}(a)]. Its shift $t_c-t_{\rm fold}$ has local exponents $0.80$, $0.91$ and $0.95$ on the ray $(-t,t)$ between successive $mL$, and it approaches $(mL)^{-1}$ from below. The slower pre-asymptotic decay reflects the compression of the condensate near the fold, which raises its compactness above the estimate~\eqref{eq:compact}. On the ray $\beta_2=0$ the fold gives $\GN M_{\max}m=0.560$, $0.605$, $0.624$ and $0.631$ for $mL=16$--$128$, against the Kaup value $0.633$. It also gives $t_{\rm fold}/\sqrt{mL}=0.650$ at $mL=128$, against $0.653$ in Eq.~\eqref{eq:kaup} [Fig.~\ref{fig:relfold}(b)]. For $\beta_2=2t$ the maximum mass keeps growing with $mL$ ($\GN M_{\max}m=1.50$ at $mL=128$), as expected for the Colpi--Shapiro--Wasserman branch.

\subsection{Column densities}

\begin{figure*}[t]
\includegraphics[width=0.85\textwidth]{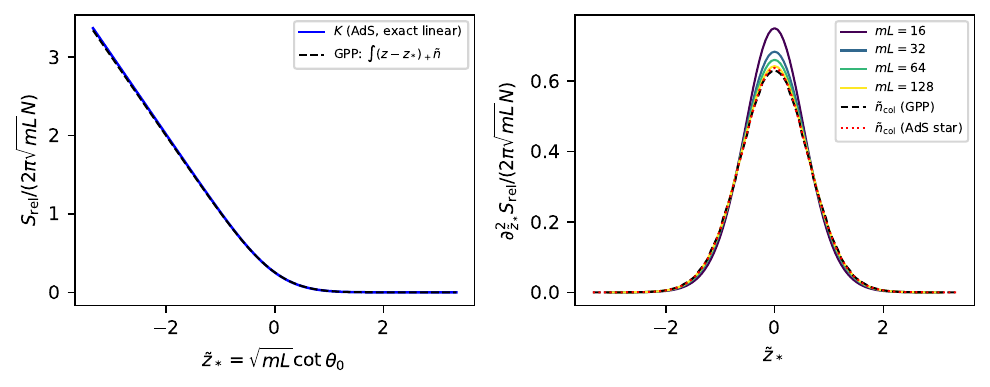}
\caption{(a) Relative entropy $K(\theta_0)$ of the AdS star (exact at linear order) and the planar transform~\eqref{eq:Fplanar} of the GPP density, against $\tilde z_*=\sqrt{mL}\cot\theta_0$. (b) $\partial^2_{\tilde z_*}K$ for increasing $mL$, compared with the GPP column density (dashed) and with the star's own column density (dotted).}
\label{fig:tomo}
\end{figure*}

Figure~\ref{fig:tomo} tests Eq.~\eqref{eq:tomo} at $t=0.3\,t_c$. Two errors separate two approximations. The first, $\|K-K_{\rm NR}[n_{\rm star}]\|$, applies the planar kernel to the star's own number density and so isolates the expansion of $\xi^t$ and $\rhoE$. The second, $\|K-K_{\rm NR}[n_{\rm GPP}]\|$, also includes the dictionary. Both decay as $(mL)^{-1.00}$, with relative errors $8.1\times10^{-3}$ and $8.0\times10^{-3}$ at $mL=128$. The kernel error dominates, so the dictionary contributes little at this order. The second derivative reproduces the star's own column density to $0.9\%$ and the GPP column density to $1.7\%$. The hemisphere entropy differs from its linearization by $0.19\%$ at $mL=128$ and by $2.4\%$ at $mL=16$.

\subsection{Three-dimensional tomography of a rotating condensate}
\label{sec:ctres}

\begin{table}[t]
\caption{Scope of the two tomography tests. $C=\GN M/r_{99}$.}
\label{tab:ctscope}
\begin{ruledtabular}
\begin{tabular}{p{1.5cm}p{2.8cm}p{2.8cm}}
 & A: rotating lattice & B: back-reacted stars \\ \hline
\tabrr{matter} & \tabrr{GPP vortex lattice, non-spherical} & \tabrr{AdS boson stars, spherical} \\
\tabrr{geometry} & \tabrr{fixed AdS} & \tabrr{fully back-reacted} \\
\tabrr{cap entropies} & \tabrr{leading order in $\GN$, Eq.~\eqref{eq:radon}} & \tabrr{exact RT areas} \\
\tabrr{inversion} & \tabrr{full angular, Eq.~\eqref{eq:inversion}} & \tabrr{radial reduction (App.~\ref{app:rt})} \\
\tabrr{measures} & \tabrr{discretization, sampling, noise, Klein vs planar kernel} & \tabrr{truncation of Eq.~\eqref{eq:radon} in $\GN$} \\
\tabrr{result} & \tabrr{on-cloud error $\le0.56\%$ for $mL\ge64$ (global $\le1.8\%$); 7 of 7 interior vortices; $\le0.11\%$ with closed-form data} & \tabrr{$\epsilon_S$ given by Eq.~\eqref{eq:epsS}; $\epsilon_{\rm img}\simeq(3.7$--$5.5)\,\epsilon_S$ below saturation} \\
\end{tabular}
\end{ruledtabular}
\end{table}

The tomography is tested twice, with complementary scope (Table~\ref{tab:ctscope}); the error measures are defined in Sec.~\ref{sec:num}. Test A checks the inversion formula on a non-spherical, rotating condensate with leading-order data on fixed AdS. Test B checks the leading-order formula itself against exact RT areas in fully back-reacted geometries, which we have only for spherical stars. Neither test covers the other.

\begin{figure*}[t]
\includegraphics[width=0.95\textwidth]{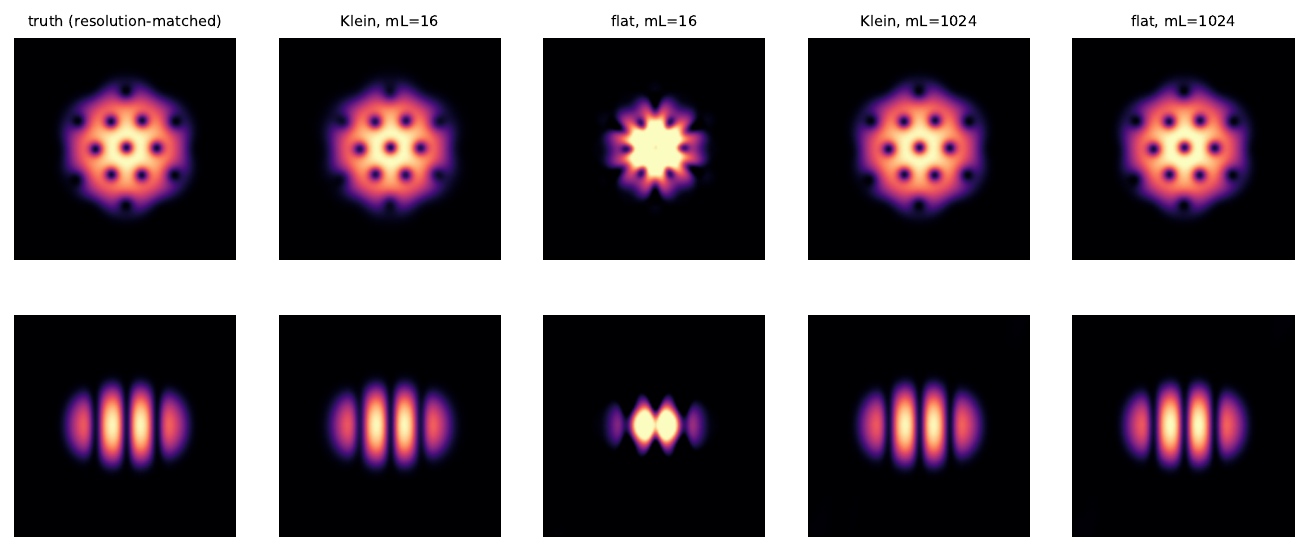}
\caption{Rotating self-gravitating condensate ($\beta_2=1000$, $\Omega_{\rm rot}=0.8\,\omega_t$, $\bG=8$) reconstructed from boundary cap entropies alone, $N_\omega=2000$. The entropies are computed at leading order in $\GN$ on fixed AdS (test A of Table~\ref{tab:ctscope}). Top row: slice $\tilde z=0$; bottom row: slice $\tilde y=0$. Left: true density blurred by the resolution kernel. Then, for $mL=16$ and $1024$: the inversion~\eqref{eq:inversion} (Klein) and its planar limit~\eqref{eq:gppct}.}
\label{fig:ct}
\end{figure*}

\begin{figure*}[t]
\includegraphics[width=0.95\textwidth]{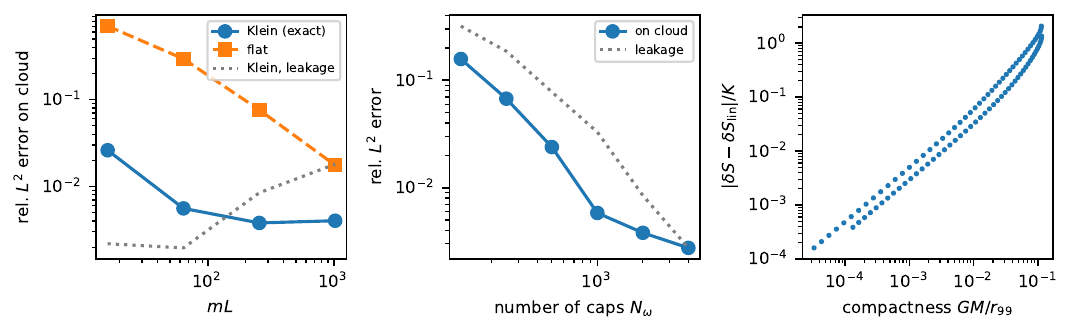}
\caption{Error budget of the tomography. (a) Relative $L^2$ error on the cloud against $mL$ for the Klein and planar inversions; dotted, out-of-cloud leakage of the Klein inversion. (b) Error against the number of cap orientations at $mL=256$. (c) Truncation error of the leading-order formula on the hemisphere, $|\delta S-\delta S_{\rm lin}|/K_{\rm hemi}$ from Eq.~\eqref{eq:Seq}, for 48 back-reacted stars each at $mL=32$ (lower band) and $128$ (upper band) on the ray $\beta_2=0$, against $C=\GN M/r_{99}$. The separation of the bands is the logarithm of Eq.~\eqref{eq:epslog}. Above the Jacobi threshold ($C\approx0.035$ at $mL=128$, above $0.047$ at $mL=32$) the symmetric disk is no longer the RT surface and the points give the disk area only.}
\label{fig:cterr}
\end{figure*}

The test state solves the rotating GPP equation~\eqref{eq:GPPdl} with $-\Omega_{\rm rot}L_z$ (Sec.~\ref{sec:gpp}) for $\beta_2=1000$, $\Omega_{\rm rot}=0.8\,\omega_t$ and $\bG=8$. It is a converged stationary state of the full equation, with the self-gravity computed from its own density at every step, not an ansatz (Sec.~\ref{sec:num}). Per particle, its rotating-frame energy is $3.940$, $\langle\tilde L_z\rangle=5.125$ and $\langle\tilde r^2\rangle=9.605$, converged to a few parts in $10^5$ (Table~\ref{tab:conv}). The same lattice without self-gravity has $5.134$, $6.200$ and $10.896$: gravity compresses the cloud, removes angular momentum, and reduces the number of vortices in the dense region from thirteen to seven. In the lab frame the lattice turns rigidly, and Eq.~\eqref{eq:inversion} images the instantaneous energy density (Sec.~\ref{sec:radon}). In full gravity such a state has only a helical Killing field. Configurations of this kind are possible in AdS because the reflecting boundary prevents the loss of gravitational radiation, as for the black holes with a single helical Killing field of Ref.~\cite{DiasHorowitzSantos2011}, but we have not constructed the back-reacted state, and test A is a test at the GPP level on fixed AdS. Thirteen vortex lines cross the plane $\tilde z=0$. Seven of them, the central line and the first hexagonal ring at $\tilde r\simeq2.2$, pass the background criterion of Sec.~\ref{sec:num}; the outer six, at $\tilde r\simeq4.0$--$4.4$, lie where the local density is $5$--$25\%$ of its maximum. We embed the condensate at $mL=16$, $64$, $256$ and $1024$, where its rms radius is $0.78$, $0.39$, $0.19$ and $0.10$ times $L$, generate the leading-order cap entropies~\eqref{eq:radon} for $N_\omega=2000$ orientations with a non-spherical $\langle T_{tt}\rangle$ ($\ell=0,2,4,6$) included, and invert them with Eq.~\eqref{eq:inversion}. Figure~\ref{fig:ct} compares the reconstruction directly with the true density. The on-cloud errors are $2.6\%$, $0.56\%$, $0.38\%$ and $0.40\%$, and the global errors, which include the leakage, are $2.6\%$, $0.59\%$, $0.93\%$ and $1.8\%$. All seven interior vortices are recovered at every $mL$. For $mL\ge64$ each lies on the same grid point as in the reference; at $mL=16$ one of the seven is displaced by one grid cell, an rms error of $0.022\,a$. The planar formula~\eqref{eq:gppct} applied to the same data has on-cloud errors of $70\%$, $29\%$, $7.6\%$ and $1.8\%$. At $mL=16$ it loses one interior vortex and the outer ring, and it compresses the cloud. The only residual of the Klein inversion that grows with $mL$ is an out-of-cloud leakage from finite angular sampling, $0.2$--$1.8\%$ of the norm, which falls with $N_\omega$ [Fig.~\ref{fig:cterr}(a)].

At $mL=256$ the on-cloud error falls from $16\%$ at $N_\omega=125$ to $0.58\%$ at $N_\omega=1000$ and $0.28\%$ at $N_\omega=4000$. The leakage falls from $32\%$ to $0.27\%$ over the same range [Fig.~\ref{fig:cterr}(b)]. For $N_\omega\ge500$ the on-cloud error scales as $N_\omega^{-1.0}$ until it reaches the grid floor, and the leakage as $N_\omega^{-1.6}$. All seven vortices are kept at every $N_\omega$ tested. Noise of relative size $\eta$ on the cap entropies is amplified by $\partial_c^4$. With a Gaussian regularization of the fourth derivative whose width minimizes the on-cloud error, the errors are $1.5\%$, $2.1\%$ and $5.0\%$ for $\eta=10^{-8}$, $10^{-7}$ and $10^{-6}$, and all seven vortices survive. Resolving vortex cores thus requires cap entropies to a relative precision of about $10^{-6}$. This is a statement about numerical holography: in the classical limit the entropies are exact.

The data of test A are binned on the same grid in $c$ that the inversion uses, which could hide errors of the inversion. We therefore repeat the test with data computed in closed form. The surrogate is a sum of Gaussians in Klein coordinates: an oblate cloud with the widths of the lattice ($\tilde\sigma_\perp=2.4$, $\tilde\sigma_z=1.4$) and seven dips elongated along $\hat z$, $90\%$ deep, at the positions of the interior vortices. Its cap integrals $\int(\bom\cdot\by-c)_+\nu\,\dd^3y$ are elementary, and the same stress tensor ($\ell=0,2,4,6$, with mean $M/4\pi$) enters through Funk--Hecke. The inversion is the production one, on the same grid in $c$ with $N_\omega=2000$, and the reference is the Gaussian sum blurred by the data kernel, evaluated exactly. On every second grid point the on-cloud errors are $0.09\%$, $0.08\%$, $0.10\%$ and $0.11\%$ at $mL=16$, $64$, $256$ and $1024$, and the leakage is $0.001\%$, $0.02\%$, $0.38\%$ and $1.3\%$. All seven dips are found at their grid positions. At $mL=256$ the on-cloud error is $4.0\%$, $0.76\%$, $0.10\%$ and $0.08\%$ for $N_\omega=500$, $1000$, $2000$ and $4000$. The stress tensor, which in these data is $2$ to $15$ times larger than the bulk term at $c=0$, changes the reconstruction by $4\times10^{-5}$ of its norm at $mL=16$ and $3\times10^{-8}$ at $mL=1024$. The inversion is thus more accurate than test A indicates. Comparing the same reconstruction with a reference blurred isotropically in $\tilde{\bm x}$, as in test A, raises the on-cloud error to $0.60\%$ at $mL=16$ and $0.25\%$ at $mL=64$; the remainder of the test-A errors comes from its discrete forward model, which bins the grid points into histograms in $c$.

These tests use cap entropies generated by the leading-order expression~\eqref{eq:radon} itself. They measure the inversion (sampling, resolution, noise, and the Klein versus planar kernel), not the accuracy of Eq.~\eqref{eq:radon}, which we test next without that assumption.

\begin{figure*}[t]
\includegraphics[width=0.95\textwidth]{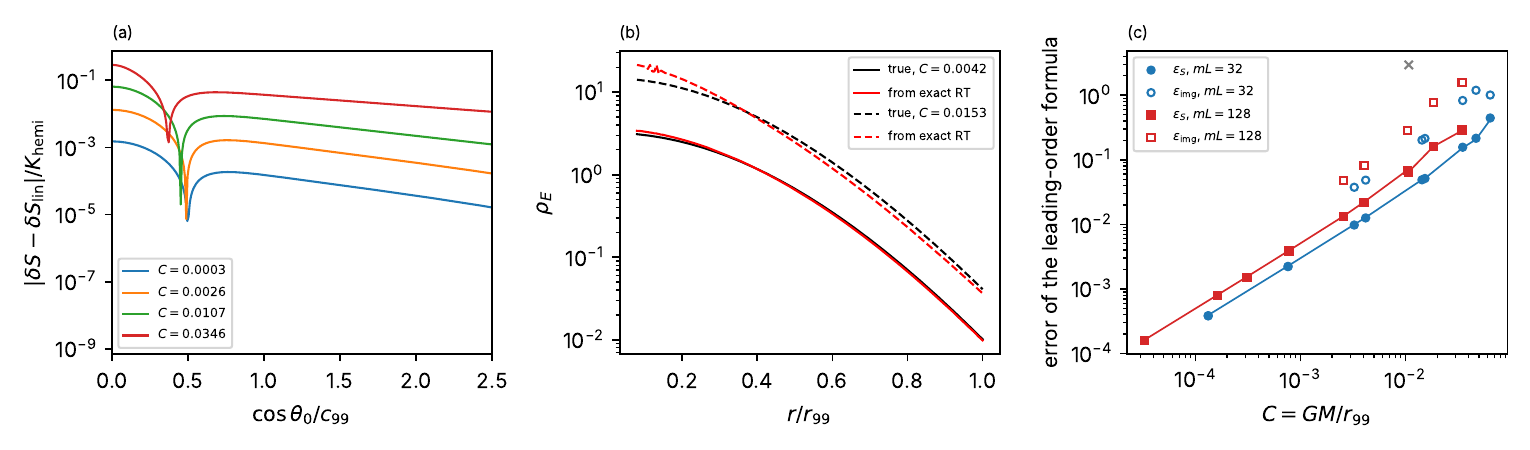}
\caption{Exact RT surfaces of all caps in back-reacted spherical stars (test B of Table~\ref{tab:ctscope}). (a) Deviation of the exact cap entropy from the leading-order one, normalized by the hemisphere $K$, against the cap position ($\beta_2=0$, $mL=128$); $c_{99}=r_{99}/\sqrt{1+r_{99}^2}$. (b) True $\rho_E$ and its reconstruction from the exact entropies by Eq.~\eqref{eq:inversion} ($\beta_2=0$, $mL=32$). (c) Error of the leading-order formula against compactness $C=\GN M/r_{99}$ for $mL=32$ (circles) and $128$ (squares). Filled: entropy error $\epsilon_S$, attained on the hemisphere; lines, the second-order formula~\eqref{eq:epsS} evaluated on each star. Open: image error $\epsilon_{\rm img}$ where it is resolved. Cross: a star whose exact-data image has an isolated spline oscillation (see text).}
\label{fig:nlrt}
\end{figure*}

We compute the exact RT surface of every cap in fully back-reacted spherical stars on the rays $\beta_2=0$ and $\beta_2=-\bG$ at $mL=32$ and $128$, six stars per ray and $mL$ on the stable branch, with compactness $C\equiv\GN M/r_{99}$ from $3\times10^{-5}$ to $0.094$ (Fig.~\ref{fig:nlrt}). The surfaces are axisymmetric minimal surfaces of the static slice, computed by shooting from the tip without subtracting divergent areas (Appendix~\ref{app:rt}). In pure AdS the construction returns $\delta S=0$ to $6\times10^{-12}$, and on the hemisphere it reproduces Eq.~\eqref{eq:Seq} to $5\times10^{-8}$. At the fold, past the point where the equatorial disk loses local minimality (Secs.~\ref{sec:hemi} and~\ref{sec:rtbound}), the cap angle is no longer a single-valued function of the tip radius. This happens for two stars, both on the ray $\beta_2=0$ where the Jacobi instability precedes the turning point ($C=0.084$ and $0.094$), and we exclude them.

For each of the remaining 22 stars the entropy error $\epsilon_S$ is attained on the hemisphere [Fig.~\ref{fig:nlrt}(a)], where Eq.~\eqref{eq:Seq} is exact. Expanding $f^{-1/2}$ in $2\GN m(r)/r$ gives, for $q=0$,
\begin{equation}
\epsilon_S=\frac{3\pi\GN}{4K_{\rm hemi}}\int_0^\infty\!\dd r\,\frac{m(r)^2}{r\,(1+r^2)^{5/2}}\,\big[1+O(C)\big].
\label{eq:epsS}
\end{equation}
The second-order term accounts for the exact remainder to $2.5\%$ for $C\le0.02$ and to $9\%$ at $C=0.064$, and Eq.~\eqref{eq:epsS} evaluated on the stored profiles reproduces the measured $\epsilon_S$ of all 22 stars to $3.5\%$ [Fig.~\ref{fig:nlrt}(c)]. For $r_{99}\ll L$ the integral is dominated by the exterior, where $m=M$. With $K_{\rm hemi}\simeq\frac{\pi}{2}M\langle r\rangle$ from Eq.~\eqref{eq:hemiNR},
\begin{equation}
\epsilon_S\simeq\frac32\,\frac{\GN M}{\langle r\rangle}\Big[\ln\frac{L}{r_{99}}+O(1)\Big].
\label{eq:epslog}
\end{equation}
The truncation error is therefore not set by the compactness alone. The RT surface feels the exterior field out to $r\sim L$, and at fixed $C$ the error grows as $\ln(L/r_{99})$. The data show this directly. For the least compact stars, $C\lesssim10^{-3}$, $\epsilon_S/C$ is $2.9$ at $mL=32$ ($r_{99}=0.43\,L$) and $4.9$ at $mL=128$ ($r_{99}=0.21\,L$); Eq.~\eqref{eq:epslog} with the Gaussian value $r_{99}/\langle r\rangle=2.11$ and the $O(1)$ term dropped gives $2.7$ and $4.9$. In the GPP regime $\GN M/\langle r\rangle=\bG/(mL\langle\tilde r\rangle)$ and $\ln(L/r_{99})=\frac12\ln(mL)+O(1)$, so the leading-order entropies converge as $(\ln mL)/mL$, more slowly by a logarithm than the dictionary of Sec.~\ref{sec:results}.

The image is more sensitive. Inverting the exact entropies overestimates the central density, by $8$--$17\%$ at $C\simeq0.0026$--$0.0042$ and by $48$--$69\%$ at $C\simeq0.010$--$0.015$, and underestimates the outer cloud [Fig.~\ref{fig:nlrt}(b)]. We call $\epsilon_{\rm img}$ resolved when it exceeds three times the $0.2$--$2.3\%$ floor of the linear pipeline and $C>2\times10^{-3}$. For eleven of the twelve resolved stars, $\epsilon_{\rm img}/\epsilon_S$ lies between $3.7$ and $5.5$ at both $mL$ and rises with $C$. At the most compact one, $C=0.064$, it is $2.3$, because the image error has saturated at order one. Combined with Eq.~\eqref{eq:epsS}, the image error is about $1$--$2\%$ at $C=10^{-3}$ and $13$--$27\%$ at $C=10^{-2}$, the larger values at $mL=128$. One further star ($\beta_2=0$, $mL=128$, $C=0.011$) gives $\epsilon_{\rm img}=2.9$, against $0.28$ for the star of equal $C$ and $mL$ on the other ray. Its error comes from an isolated oscillation of the spline derivative at $r\simeq0.28\,r_{99}$, which points to a single inaccurate cap; elsewhere its image follows the other stars, and we exclude it.

As Table~\ref{tab:ctscope} shows, test A exercises the full angular inversion, but with leading-order data. Test B supplies exact data, but for radial $\rhoE$ the inversion reduces to the one-dimensional transform $\nu(c)=\partial_c^3[s\,\delta S/2\pi]/(2\pi c)$ of Appendix~\ref{app:rt}, and the angular back-projection is not tested beyond linear order. Exact RT surfaces in a back-reacted, non-spherical geometry were not computed, so the truncation error of the rotating image is an estimate, not a measurement. Equation~\eqref{eq:epsS} gives the entropy error at second order. We take $m(r)$ from the angular average of the cloud and $K$ for the hemisphere normal to the rotation axis, which has the smallest $K$ because the cloud is oblate ($\langle|\tilde z|\rangle=1.09$, $\langle\tilde r\rangle=2.94$). This gives $\epsilon_S\simeq4.7\%$, $3.9\%$, $2.1\%$ and $0.9\%$ at $mL=16$, $64$, $256$ and $1024$. With the image-to-entropy ratio of four to five found for spherical stars, the truncation error of the rotating image would be about $20\%$, $17\%$, $9\%$ and $4\%$, larger than the inversion error at every $mL$ tested. Smaller $\bG$ or larger $mL$ moves the balance to the inversion. The Klein-versus-planar comparison is unaffected, because it concerns the kernel and holds for any $\rhoE$ at small $\GN$.

\subsection{Fold susceptibility and the RT bound}
\label{sec:rtbound}

\begin{figure*}[t]
\includegraphics[width=0.85\textwidth]{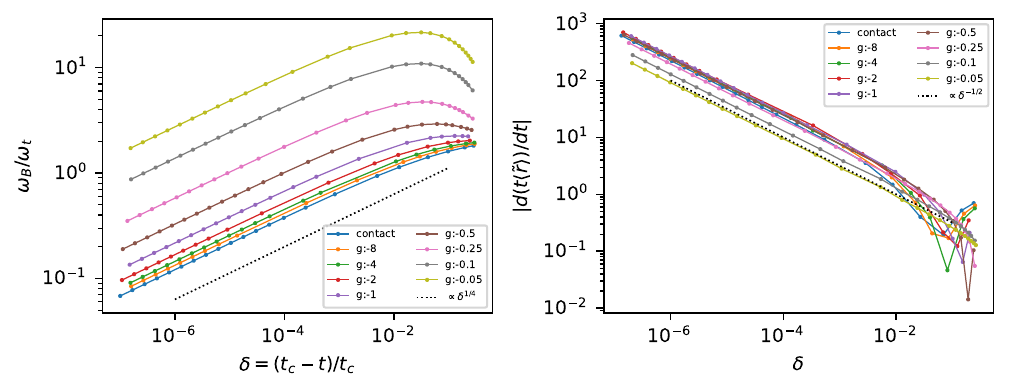}
\caption{(a) Breathing frequency against $\delta=(t_c-t)/t_c$ on all rays with a Newtonian fold; dotted line, $\delta^{1/4}$. (b) Entanglement susceptibility $|\chi_E|$; dotted line, $\delta^{-1/2}$.}
\label{fig:fold}
\end{figure*}

The susceptibility diverges with exponents between $-0.494$ and $-0.518$ (mean $-0.500$), fitted within $\delta<3\times10^{-3}$ of each fold, against the predicted $-1/2$ (Fig.~\ref{fig:fold}). The product $\chi_E\tilde\omega_B^2$ tends to the amplitudes $\mathcal A$ of Table~\ref{tab:folds}.

The RT surface itself behaves differently for the two kinds of collapse. On rays where collapse is relativistic ($\beta_2=0$ and $\beta_2=2t$), the Jacobi operator of the equatorial disk acquires a negative mode \emph{before} the turning point at every $mL$. The ratio $\psi_0^{J}/\psi_0^{\rm fold}$ falls from $0.93$ to $0.57$ ($\beta_2=0$) and from $0.92$ to $0.64$ ($\beta_2=2t$) as $mL$ goes from $16$ to $128$. On the ray $\beta_2=2t$ at $mL=128$, where $2\GN M/r_*\simeq0.10$ at the fold with $r_*$ the outer edge of the integrated star, even the positivity bound is violated on the stable branch, at $\psi_0=0.181$ against $\psi_0^{\rm fold}=0.202$. On the neutral attractive ray, whose fold is Newtonian, the disk stays locally minimal through the fold for $mL\ge32$. The RT surface therefore breaks reflection symmetry while the star is still dynamically stable when collapse is relativistic, but not when it is driven by contact attraction. The charged sequence interpolates: the instability precedes the fold for $mL\le32$ and follows it for $mL\ge64$.

The loss of minimality has three consequences. First, it is a reorganization of the entanglement of the dual state that precedes its dynamical instability. Beyond the threshold the hemisphere entropy is computed by one of a pair of mirror-image surfaces, which have equal areas by reflection symmetry. Along the sequence of stars the entropy therefore stays continuous, but the area difference between the symmetric and the asymmetric branches grows quadratically past a pitchfork, so we expect a jump in the second derivative of $\delta S_{\rm hemi}$ along the sequence. We have not computed the asymmetric branch, and this remains an expectation. In this sense the boundary entanglement registers the approach of relativistic collapse before the breathing mode softens, whereas for collapse driven by contact attraction the only signal is the susceptibility of Sec.~\ref{sec:fold}. Second, it ends the domain of the leading-order tomography. The expansion in $\GN$ follows the reflection-symmetric extremal surface, which beyond the threshold is not the RT surface of near-equatorial caps. For a fixed star, the minimal surfaces of caps with $c>0$ and $c<0$ then belong to different branches, and $\delta S(c)$ can have a kink at $c=0$, which $\partial_c^4$ would turn into a singular artifact on every plane through the center. The non-monotone map $r_{\rm tip}\mapsto c$ in the two excluded stars of test B is the numerical sign of these competing surfaces. Third, the limitation is weak in practice. On the ray $\beta_2=0$ of test B the threshold lies near $C=0.035$ at $mL=128$, where the star of that compactness sits at $\psi_0/\psi_0^J=1.02$ and still has a single-valued map, and above $C=0.047$ at $mL=32$. There the leading-order entropies are already in error by $29\%$ and by at least $22\%$, so the image has ceased to be quantitative before minimality is lost. A condensate in the GPP regime has $C=O(\bG/mL)$ and is far from the threshold.

% ======================================================================
\section{Discussion}
\label{sec:concl}

The holographic route to the GPP equation is more than another derivation of it. AdS removes the arbitrariness of the trap: the frequency is fixed by the cosmological constant, and the anharmonic and relativistic corrections are all $O(1/mL)$. The route also attaches to the condensate a dual quantum state, and the effects found here live in that description. At the fold the entanglement susceptibility of the dual state diverges, with the whole singular response in the bulk relative entropy. The entanglement entropies of caps image the condensate in three dimensions and suffice on their own: the Klein structure turns them into Radon data in which the boundary stress tensor is a shell on the boundary of the ball, and the inversion removes it from the interior, an explicit and tested instance of the kinematic-space separation of boundary and bulk contributions. The error of the leading-order image is a second-order effect of the exterior gravitational field, of size $(\GN M/\langle r\rangle)\ln(L/r_{99})$, so in the GPP regime the tomography converges as $(\ln mL)/mL$. When collapse is relativistic, the Jacobi instability of the RT disk precedes the dynamical instability of the star. The many-body, Hartree and $1/c$ routes produce the same equation without a dual state, and none of these effects has a counterpart there.

Several extensions follow. The rotating states of Sec.~\ref{sec:ctres} were imaged on a fixed AdS background, which is consistent at the order of Eq.~\eqref{eq:inversion}. Their back-reaction is a Kerr--AdS-like, non-spherical geometry~\cite{HawkingHunterTaylorRobinson1999}, and its RT surfaces would turn the estimate of the truncation error in Sec.~\ref{sec:ctres} into a measurement for non-spherical matter; this requires an axisymmetric or three-dimensional Einstein solver and is the main open test of the tomography. The $O(N^0)$ bulk entanglement across the RT plane is the first quantum correction to Eq.~\eqref{eq:inversion}~\cite{FaulknerLewkowyczMaldacena2013}. It is sensitive to quantum depletion and to vortex-core structure, and whether it resolves depletion near the fold is open. The competing RT surfaces that take over beyond the Jacobi instability of the equatorial disk need a minimal-surface solver without reflection symmetry, and real-time collapse needs the covariant HRT prescription~\cite{HRT2007}.

\paragraph*{Data availability.}
The data and code that support the findings of this article are available from the author upon reasonable request.

\begin{acknowledgments}
The author thanks Dinh-Thi Nguyen for introducing him to the Gross–Pitaevskii–Poisson system, and for guidance and discussions. AI tools (Claude, Anthropic) assisted in drafting the text and the code; the author checked all derivations and results and takes full responsibility for the content.
\end{acknowledgments}

\appendix

\section{Order counting in the reduction}
\label{app:nr}

Let $\epsilon=(mL)^{-1}$ and use trap variables $r=\ell\tilde r$, $\omega-m=\tilde\mu/L$. Each relativistic term in Eq.~\eqref{eq:KG} is then a power of $\epsilon$ times a trap-scale quantity. First, $e^\chi\Omega^2/f^2-U/f=2m(\delta\omega+q\phi)-2m^2(r^2/2L^2+\Phi_N)-\lambda\psi^2-\gamma\psi^4+R_1$, where $R_1$ collects $(\delta\omega+q\phi)^2$, $m^2\Phi_N^2$, $m^2r^4/L^4$ and the cross terms. Relative to the retained terms, $R_1/(2m\delta\omega)=O(\epsilon)$. Second, $f'/f-\chi'/2=2r/L^2+O(\Phi_N/r)$ modifies the Laplacian at relative order $\ell^2/L^2=\epsilon$. Third, with $\psi=\varphi/\sqrt{2m}$ and $|\varphi|^2=N\tilde n/\ell^3$, one has $\lambda\psi^2/m^2=2\beta_2\tilde n\,\epsilon$ and $\gamma\psi^4/m^2=2\beta_3\tilde n^2\epsilon$. Fourth, the sources in Eqs.~\eqref{eq:Max}--\eqref{eq:f} are $2qm\psi^2[1+O(\epsilon)]$ and $2m^2\psi^2[1+O(\epsilon)]$. At fixed $\beta$'s the reduced system is therefore Eq.~\eqref{eq:GPP} with $O(\epsilon)$ errors, uniformly in $\tilde r$ on the support of $\tilde\varphi$. The estimate degrades near a fold, where $\tilde\varphi$ is compressed and the effective expansion parameter grows.

\section{Exterior contributions}
\label{app:tails}

For $r>r_*$ the energy density is purely electric, $\rhoE=q^2N^2/32\pi^2r^4$, and $m(r)=M-q^2N^2/8\pi r$. The tails added to the interior integrals are
\begin{align*}
\int_{r_*}^\infty\!\frac{m(r)\,\dd r}{(1+r^2)^{3/2}}&=M\Big(1-\frac{r_*}{\sqrt{1+r_*^2}}\Big)\\
&\quad-\frac{q^2N^2}{8\pi}\Big(\operatorname{arsinh}\frac1{r_*}-\frac{1}{\sqrt{1+r_*^2}}\Big),\\
\int_{r_*}^\infty\!\frac{r^3\rhoE\,\dd r}{\sqrt{1+r^2}}&=\frac{q^2N^2}{32\pi^2}\operatorname{arsinh}\frac1{r_*},
\end{align*}
and the exterior part of Eq.~\eqref{eq:Seq} is evaluated by Gauss--Legendre quadrature in $1/r$, using Eq.~\eqref{eq:MN} and the cancellation-free form $f^{-1/2}-f_0^{-1/2}=(f_0-f)/[\sqrt{ff_0}(\sqrt f+\sqrt{f_0})]$. The physical frequency is $\omega=[\Omega(r_*)+q^2N e^{-\chi_*/2}/4\pi r_*]\,e^{\chi_*/2}$. This restores $\chi(\infty)=0$ from the integration frame $\chi(0)=0$.

\section{Klein coordinates and the boundary term}
\label{app:klein}

Global $\AdS_4$ ($L=1$) is the hyperboloid $-X_{-1}^2-X_0^2+\bm X^2=-1$ with $X_0=\sqrt{1+r^2}\cos t$, $X_{-1}=\sqrt{1+r^2}\sin t$ and $\bm X=\bm x$. The wedge of the cap $\bom\cdot\bm n>c_0$ is preserved by
\begin{multline*}
\xi=\frac{2\pi}{s_0}\Big[(\bom\cdot\bm X)\partial_{X_{-1}}+X_{-1}(\bom\cdot\partial_{\bm X})\Big]\\
-\frac{2\pi c_0}{s_0}\Big[X_0\partial_{X_{-1}}-X_{-1}\partial_{X_0}\Big],
\end{multline*}
which vanishes on the RT surface, has surface gravity $2\pi$, and reduces on the boundary to $2\pi(\bom\cdot\bm n-c_0)s_0^{-1}\partial_t$~\cite{CasiniHuertaMyers2011}. At $t=0$, $\partial_{X_{-1}}=(1+r^2)^{-1/2}\partial_t$, which gives Eq.~\eqref{eq:xiklein} with $\by=\bm X/X_0$, the Beltrami--Klein projection. On the slice, $n=(1+r^2)^{-1/2}\partial_t$, $\dd\Sigma=(1+r^2)^{-1/2}r^2\dd r\,\dd\Omega$ and $T_{tt}=(1+r^2)\rhoE$, so $\xi^\mu T_{\mu\nu}n^\nu\dd\Sigma=\xi^t\rhoE\,\dd^3x$. With $\dd|\by|/\dd r=(1+r^2)^{-3/2}$ and $|\by|^2/r^2=(1+r^2)^{-1}$ one finds $\dd^3x=(1-|\by|^2)^{-5/2}\dd^3y$. The metric perturbation enters $K$ only at $O(\GN^2)$.

At $t=0$, where $X_{-1}=0$, the vector $\xi$ is proportional to $\partial_{X_{-1}}\propto\partial_t$ and hence normal to the slice. For a rotating state the pattern turns rigidly in the lab frame and the bulk is stationary but not static. Caps anchored at boundary time $t$ have their RT surfaces on the global slice $t$ at this order, and the time-translated $\xi$ is again normal there. Only $\xi^tT_{tt}$ enters $K$; the momentum density of the rotating condensate does not, and Eq.~\eqref{eq:inversion} images the instantaneous energy density.

The boundary shell drops out at interior points, as stated in Sec.~\ref{sec:radon}. To see it explicitly, expand $\langle T_{tt}\rangle=\sum_\ell T_\ell P_\ell(\bm n\cdot\bm e)$; every spherical harmonic of degree $\ell$ is a finite sum of such zonal terms with different axes $\bm e$. By the Funk--Hecke theorem, $\int_{S^2}\delta(\bom\cdot\bm n-c)P_\ell(\bm n\cdot\bm e)\,\dd\Omega=2\pi P_\ell(c)P_\ell(\bom\cdot\bm e)$ for $|c|<1$. The shell's contribution to $\partial_c^2$ of the Radon data is $2\pi T_\ell P_\ell''(c)P_\ell(\bom\cdot\bm e)$. At $c=\bom\cdot\by$ with $|\by|<1$, $P_\ell''(\bom\cdot\by)$ is a polynomial of degree $\ell-2$ in the components of $\bom$ and so contains only spherical harmonics of degree $\le\ell-2$. Its integral against $P_\ell(\bom\cdot\bm e)$ over $S^2$ vanishes. The modular energy thus drops out of Eq.~\eqref{eq:inversion} at every interior point, for any $\langle T_{tt}\rangle$.

\section{Exact RT surfaces of spherical stars}
\label{app:rt}

On the static slice $\dd r^2/f+r^2\dd\Omega^2$ of a back-reacted star, the RT surface of the cap $\theta<\theta_0$ is axisymmetric. Its area is $A=2\pi\int r\sin\theta\,(r^2\dd\theta^2+\dd r^2/f)^{1/2}$. Near the axis we parametrize it as $r(\theta)$. The Euler--Lagrange equation is regular at $\theta=0$, with $r(0)=r_{\rm tip}$, $r'(0)=0$ and $r''(0)=r_{\rm tip}f(r_{\rm tip})$. Once $r'>4r$ or $r>\max(2,3r_{\rm tip})$ we switch to $\theta(r)$ with independent variable $\ln r$ and integrate to $R=10^6$; $f'$ is taken from Eq.~\eqref{eq:f}, so the interior profile enters only through $f$ and $\rhoE$. Instead of the divergent area we integrate
\begin{equation}
D=\frac{A}{2\pi}-\Big(\sqrt{1+r^2}\,\sin\theta-1\Big),
\end{equation}
whose derivative falls off as $r^{-2}$. Every RT surface of pure AdS has $D(\infty)=0$: it is a totally geodesic plane at hyperbolic distance $\operatorname{artanh}c$ from the origin, whose area inside $r<R$ is $2\pi(s\sqrt{1+R^2}-1)$. Therefore $\delta S=(\pi/2\GN)D(\infty)$ for the star, with $D(\infty)=D(R)-c^2/(2sR)+O(R^{-2})$. At large $r$ the integrand is written as $r\sin\theta\,(r^2\theta_r^2+(f_0-f)/ff_0)/(V+f_0^{-1/2})-\sqrt{f_0}\cos\theta\,\theta_r$, with $V=(r^2\theta_r^2+1/f)^{1/2}$ and $f_0=1+r^2$, so that no cancellation occurs. The cap is read off as $c=\cos\theta(R)\,R/\sqrt{1+R^2}$, exact in vacuum and correct to $O(\GN M/R^3)$ otherwise. We integrate with an eighth-order Runge--Kutta method at relative tolerance $10^{-12}$. For a star with tip radii chosen on a grid in $c$, the map $r_{\rm tip}\mapsto c$ is single-valued away from the Jacobi instability, and $\delta S(c)$ is even in $c$ by purity. For radial data Eq.~\eqref{eq:inversion} reduces to $\nu(c)=\partial_c^3[s\,\delta S/2\pi]/(2\pi c)$, which we evaluate with a quintic interpolating spline.

\section{Self-bound branch}
\label{sec:droplet}

For $\beta_2<0<\beta_3$ the sextic term arrests collapse. The branch $\tilde\mu(t)$ then becomes S-shaped, with a dilute, trap-dominated state and a dense, flat-topped droplet whose density approaches $\tilde n_*=3|\beta_2|/4\beta_3$. This is the trapped analogue of quantum droplets~\cite{Petrov2015} and, in the relativistic theory, of gravitating $Q$-balls~\cite{Coleman1985,KleihausKunzList2005}. At fixed $(\beta_3,\bG)$ the two spinodals $\beta_2^{\rm dil}$ and $\beta_2^{\rm den}$ are folds of the type of Sec.~\ref{sec:fold}, and the first-order (Maxwell) line lies between them. As $\beta_3$ grows, the spinodals merge at a cusp $(\beta_{3c},\beta_{2c})$ with gap $\beta_2^{\rm den}-\beta_2^{\rm dil}\propto(\beta_{3c}-\beta_3)^{3/2}$. The cusp is a liquid--gas critical end point, not a tricritical point: the problem has no symmetry that a tricritical point would require. Self-gravity shifts the cusp.

\begin{figure}[b]
\includegraphics[width=\columnwidth]{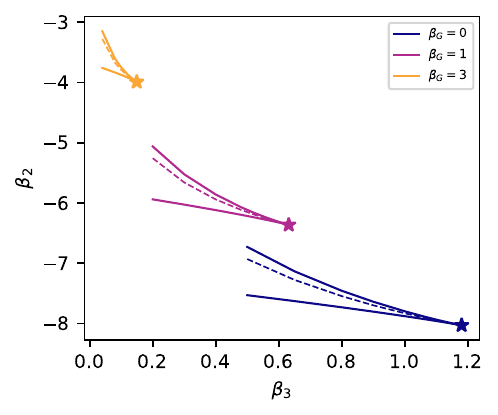}
\caption{Spinodals (solid) and Maxwell line (dashed) of the dilute--droplet transition for $\bG=0,1,3$; stars mark the cusps.}
\label{fig:drop}
\end{figure}

Both spinodals are tracked by continuation, with a continuity condition on $\langle\tilde r^2\rangle$ that prevents the solver from jumping between branches near the cusp. The cusp is located by fitting the spinodal gap to the $(\beta_{3c}-\beta_3)^{3/2}$ law on the four bistable points closest to it. This gives $(\beta_{3c},\beta_{2c})=(1.180,-8.030)$ for $\bG=0$, $(0.632,-6.364)$ for $\bG=1$ and $(0.149,-3.993)$ for $\bG=3$ (Fig.~\ref{fig:drop}); each cusp lies inside the last bistable interval of the computed spinodals. Self-gravity narrows the coexistence region and moves the cusp to smaller three-body coupling. The dilute branch is destabilized at weaker attraction, while the dense droplet is pulled together by gravity before $\beta_3$ can stabilize it.

\section{Convergence and error budget}
\label{app:conv}

\begin{table*}[t]
\caption{Grid convergence of the rotating state ($\beta_2=1000$, $\Omega_{\rm rot}=0.8\,\omega_t$, $\bG=8$). Differences are relative to the $192^3$ grid on $[-8,8)^3$; $\delta n$ is the relative $L^2$ difference of the density in the plane $\tilde z=0$; $\delta x_v$ is the rms vortex displacement in lattice spacings; $\epsilon_{\rm CT}$ is the on-cloud tomography error at $mL=256$.}
\label{tab:conv}
\begin{ruledtabular}
\begin{tabular}{lcccccc}
grid & $\delta E$ & $\delta L_z$ & $\delta\langle\tilde r^2\rangle$ & $\delta n$ & $\delta x_v$ & $\epsilon_{\rm CT}$ \\ \hline
$96^3$ & $3.7\times10^{-5}$ & $8.4\times10^{-5}$ & $5.0\times10^{-5}$ & $3.0\times10^{-4}$ & $0.044$ & $1.35\%$ \\
$128^3$ & $1.5\times10^{-5}$ & $3.5\times10^{-5}$ & $2.1\times10^{-5}$ & $7.8\times10^{-5}$ & $0.030$ & $0.38\%$ \\
$160^3$ & $5.4\times10^{-6}$ & $1.2\times10^{-5}$ & $7.4\times10^{-6}$ & $2.6\times10^{-5}$ & $0.023$ & $0.22\%$ \\
$192^3$ & --- & --- & --- & --- & --- & $0.16\%$ \\
$160^3$, $[-10,10)^3$ & $1.5\times10^{-5}$ & $3.5\times10^{-5}$ & $2.1\times10^{-5}$ & $7.8\times10^{-5}$ & $0.030$ & $0.38\%$ \\
\end{tabular}
\end{ruledtabular}
\end{table*}

Table~\ref{tab:conv} lists the grid study of the rotating state. Successive absolute energy differences, $8.5\times10^{-5}$, $3.9\times10^{-5}$ and $2.1\times10^{-5}$, are in the ratios $2.16$ and $1.84$ expected for second-order convergence on this sequence of $\Delta x$, the order of the Hockney kernel. The enlarged box at the production spacing reproduces the $128^3$ values to nine digits, so the box is large enough. The tomography error falls with $\Delta x$, which shows that the inversion errors of Sec.~\ref{sec:ctres} are discretization errors.

The error of the reconstruction has four sources. Discretization ($\Delta x$, $\Delta c$, the five-point $\partial_c^4$) is measured by the Gaussian round trip of Table~\ref{tab:gates} and by Table~\ref{tab:conv}. Angular sampling gives errors that scale as $N_\omega^{-1}$ on the cloud and $N_\omega^{-1.6}$ outside it. Noise on the cap entropies is amplified by $\partial_c^4$, roughly as $\eta\,(\Delta c)^{-4}$ before regularization. Truncation at leading order in $\GN$ is measured with exact RT surfaces (Fig.~\ref{fig:nlrt}). For the fold amplitudes, the estimates from the three windows, the two polynomial degrees and the two radial grids agree to $2\times10^{-7}$.

\bibliography{Holo_BEC}

@article{Kaup1968,
  author  = {Kaup, David J.},
  title   = {{Klein-Gordon Geon}},
  journal = {Phys. Rev.},
  volume  = {172}, pages = {1331}, year = {1968},
  doi     = {10.1103/PhysRev.172.1331}
}

@article{RuffiniBonazzola1969,
  author  = {Ruffini, Remo and Bonazzola, Silvano},
  title   = {{Systems of Self-Gravitating Particles in General Relativity and the Concept of an Equation of State}},
  journal = {Phys. Rev.},
  volume  = {187}, pages = {1767}, year = {1969},
  doi     = {10.1103/PhysRev.187.1767}
}

@article{Colpi1986,
  author  = {Colpi, Monica and Shapiro, Stuart L. and Wasserman, Ira},
  title   = {{Boson Stars: Gravitational Equilibria of Self-Interacting Scalar Fields}},
  journal = {Phys. Rev. Lett.},
  volume  = {57}, pages = {2485}, year = {1986},
  doi     = {10.1103/PhysRevLett.57.2485}
}

@article{JetzerVanDerBij1989,
  author  = {Jetzer, Philippe and van der Bij, J. J.},
  title   = {{Charged boson stars}},
  journal = {Phys. Lett. B},
  volume  = {227}, pages = {341}, year = {1989},
  doi     = {10.1016/0370-2693(89)90941-6}
}

@article{LeePang1989,
  author  = {Lee, T. D. and Pang, Y.},
  title   = {{Stability of mini-boson stars}},
  journal = {Nucl. Phys. B},
  volume  = {315}, pages = {477}, year = {1989},
  doi     = {10.1016/0550-3213(89)90365-4}
}

@article{GleiserWatkins1989,
  author  = {Gleiser, Marcelo and Watkins, Richard},
  title   = {{Gravitational stability of scalar matter}},
  journal = {Nucl. Phys. B},
  volume  = {319}, pages = {733}, year = {1989},
  doi     = {10.1016/0550-3213(89)90627-5}
}

@article{Kusmartsev1991,
  author  = {Kusmartsev, F. V. and Mielke, E. W. and Schunck, F. E.},
  title   = {{Gravitational stability of boson stars}},
  journal = {Phys. Rev. D},
  volume  = {43}, pages = {3895}, year = {1991},
  doi     = {10.1103/PhysRevD.43.3895}
}

@article{SchunckMielke2003,
  author  = {Schunck, Franz E. and Mielke, Eckehard W.},
  title   = {{General relativistic boson stars}},
  journal = {Class. Quantum Grav.},
  volume  = {20}, pages = {R301}, year = {2003},
  doi     = {10.1088/0264-9381/20/20/201}
}

@article{LieblingPalenzuela2017,
  author  = {Liebling, Steven L. and Palenzuela, Carlos},
  title   = {{Dynamical boson stars}},
  journal = {Living Rev. Relativ.},
  volume  = {20}, pages = {5}, year = {2017},
  doi     = {10.1007/s41114-017-0007-y}
}

@article{Chavanis2011,
  author  = {Chavanis, Pierre-Henri},
  title   = {{Mass-radius relation of Newtonian self-gravitating Bose-Einstein condensates with short-range interactions. I. Analytical results}},
  journal = {Phys. Rev. D},
  volume  = {84}, pages = {043531}, year = {2011},
  doi     = {10.1103/PhysRevD.84.043531}
}

@article{ChavanisDelfini2011,
  author  = {Chavanis, Pierre-Henri and Delfini, L.},
  title   = {{Mass-radius relation of Newtonian self-gravitating Bose-Einstein condensates with short-range interactions. II. Numerical results}},
  journal = {Phys. Rev. D},
  volume  = {84}, pages = {043532}, year = {2011},
  doi     = {10.1103/PhysRevD.84.043532}
}

@article{Hui2017,
  author  = {Hui, Lam and Ostriker, Jeremiah P. and Tremaine, Scott and Witten, Edward},
  title   = {{Ultralight scalars as cosmological dark matter}},
  journal = {Phys. Rev. D},
  volume  = {95}, pages = {043541}, year = {2017},
  doi     = {10.1103/PhysRevD.95.043541}
}

@article{ODell2000,
  author  = {O'Dell, D. and Giovanazzi, S. and Kurizki, G. and Akulin, V. M.},
  title   = {{Bose-Einstein Condensates with $1/r$ Interatomic Attraction: Electromagnetically Induced ``Gravity''}},
  journal = {Phys. Rev. Lett.},
  volume  = {84}, pages = {5687}, year = {2000},
  doi     = {10.1103/PhysRevLett.84.5687}
}

@article{Petrov2015,
  author  = {Petrov, D. S.},
  title   = {{Quantum Mechanical Stabilization of a Collapsing Bose-Bose Mixture}},
  journal = {Phys. Rev. Lett.},
  volume  = {115}, pages = {155302}, year = {2015},
  doi     = {10.1103/PhysRevLett.115.155302}
}

@article{Maldacena1998,
  author  = {Maldacena, Juan M.},
  title   = {{The Large $N$ limit of superconformal field theories and supergravity}},
  journal = {Adv. Theor. Math. Phys.},
  volume  = {2}, pages = {231}, year = {1998},
  doi     = {10.4310/ATMP.1998.v2.n2.a1}
}

@article{HHH2008b,
  author  = {Hartnoll, Sean A. and Herzog, Christopher P. and Horowitz, Gary T.},
  title   = {{Holographic superconductors}},
  journal = {J. High Energy Phys.},
  volume  = {12}, pages = {015}, year = {2008},
  doi     = {10.1088/1126-6708/2008/12/015}
}

@article{AstefaneseiRadu2003,
  author  = {Astefanesei, Dumitru and Radu, Eugen},
  title   = {{Boson stars with negative cosmological constant}},
  journal = {Nucl. Phys. B},
  volume  = {665}, pages = {594}, year = {2003},
  doi     = {10.1016/S0550-3213(03)00481-5}
}

@article{Basu2010,
  author  = {Basu, Pallab and Bhattacharya, Jyotirmoy and Bhattacharyya, Sayantani and Loganayagam, R. and Minwalla, Shiraz and Umesh, V.},
  title   = {{Small hairy black holes in global AdS spacetime}},
  journal = {J. High Energy Phys.},
  volume  = {10}, pages = {045}, year = {2010},
  doi     = {10.1007/JHEP10(2010)045}
}

@article{Dias2012,
  author  = {Dias, {\'O}scar J. C. and Horowitz, Gary T. and Marolf, Donald and Santos, Jorge E.},
  title   = {{On the nonlinear stability of asymptotically anti-de Sitter solutions}},
  journal = {Class. Quantum Grav.},
  volume  = {29}, pages = {235019}, year = {2012},
  doi     = {10.1088/0264-9381/29/23/235019}
}

@article{BuchelLieblingLehner2013,
  author  = {Buchel, Alex and Liebling, Steven L. and Lehner, Luis},
  title   = {{Boson stars in AdS spacetime}},
  journal = {Phys. Rev. D},
  volume  = {87}, pages = {123006}, year = {2013},
  doi     = {10.1103/PhysRevD.87.123006}
}

@article{BizonEvninFicek2018,
  author  = {Bizo{\'n}, Piotr and Evnin, Oleg and Ficek, Filip},
  title   = {{A nonrelativistic limit for AdS perturbations}},
  journal = {J. High Energy Phys.},
  volume  = {12}, pages = {113}, year = {2018},
  doi     = {10.1007/JHEP12(2018)113}
}

@article{Ficek2021,
  author  = {Ficek, Filip},
  title   = {{Schr\"odinger-Newton-Hooke system in higher dimensions: Stationary states}},
  journal = {Phys. Rev. D},
  volume  = {103}, pages = {104062}, year = {2021},
  doi     = {10.1103/PhysRevD.103.104062}
}

@article{BreitenlohnerFreedman1982,
  author  = {Breitenlohner, Peter and Freedman, Daniel Z.},
  title   = {{Stability in gauged extended supergravity}},
  journal = {Ann. Phys. (N.Y.)},
  volume  = {144}, pages = {249}, year = {1982},
  doi     = {10.1016/0003-4916(82)90116-6}
}

@article{RyuTakayanagi2006,
  author  = {Ryu, Shinsei and Takayanagi, Tadashi},
  title   = {{Holographic Derivation of Entanglement Entropy from the anti-de Sitter Space/Conformal Field Theory Correspondence}},
  journal = {Phys. Rev. Lett.},
  volume  = {96}, pages = {181602}, year = {2006},
  doi     = {10.1103/PhysRevLett.96.181602}
}

@article{HRT2007,
  author  = {Hubeny, Veronika E. and Rangamani, Mukund and Takayanagi, Tadashi},
  title   = {{A covariant holographic entanglement entropy proposal}},
  journal = {J. High Energy Phys.},
  volume  = {07}, pages = {062}, year = {2007},
  doi     = {10.1088/1126-6708/2007/07/062}
}

@article{Blanco2013,
  author  = {Blanco, David D. and Casini, Horacio and Hung, Ling-Yan and Myers, Robert C.},
  title   = {{Relative entropy and holography}},
  journal = {J. High Energy Phys.},
  volume  = {08}, pages = {060}, year = {2013},
  doi     = {10.1007/JHEP08(2013)060}
}

@article{Faulkner2014,
  author  = {Faulkner, Thomas and Guica, Monica and Hartman, Thomas and Myers, Robert C. and Van Raamsdonk, Mark},
  title   = {{Gravitation from entanglement in holographic CFTs}},
  journal = {J. High Energy Phys.},
  volume  = {03}, pages = {051}, year = {2014},
  doi     = {10.1007/JHEP03(2014)051}
}

@article{JLMS2016,
  author  = {Jafferis, Daniel L. and Lewkowycz, Aitor and Maldacena, Juan and Suh, S. Josephine},
  title   = {{Relative entropy equals bulk relative entropy}},
  journal = {J. High Energy Phys.},
  volume  = {06}, pages = {004}, year = {2016},
  doi     = {10.1007/JHEP06(2016)004}
}

@article{FaulknerLewkowyczMaldacena2013,
  author  = {Faulkner, Thomas and Lewkowycz, Aitor and Maldacena, Juan},
  title   = {{Quantum corrections to holographic entanglement entropy}},
  journal = {J. High Energy Phys.},
  volume  = {11}, pages = {074}, year = {2013},
  doi     = {10.1007/JHEP11(2013)074}
}

@article{Lin2015,
  author  = {Lin, Jennifer and Marcolli, Matilde and Ooguri, Hirosi and Stoica, Bogdan},
  title   = {{Locality of Gravitational Systems from Entanglement of Conformal Field Theories}},
  journal = {Phys. Rev. Lett.},
  volume  = {114}, pages = {221601}, year = {2015},
  doi     = {10.1103/PhysRevLett.114.221601}
}

@article{Lashkari2016,
  author  = {Lashkari, Nima and Lin, Jennifer and Ooguri, Hirosi and Stoica, Bogdan and Van Raamsdonk, Mark},
  title   = {{Gravitational positive energy theorems from information inequalities}},
  journal = {Prog. Theor. Exp. Phys.}, volume = {2016}, pages = {12C109}, year = {2016},
  doi     = {10.1093/ptep/ptw139},
  eprint  = {1605.01075}
}

@article{Ruprecht1995,
  author  = {Ruprecht, P. A. and Holland, M. J. and Burnett, K. and Edwards, Mark},
  title   = {{Time-dependent solution of the nonlinear Schr\"odinger equation for Bose-condensed trapped neutral atoms}},
  journal = {Phys. Rev. A},
  volume  = {51}, pages = {4704}, year = {1995},
  doi     = {10.1103/PhysRevA.51.4704}
}

@article{UedaLeggett1998,
  author  = {Ueda, Masahito and Leggett, Anthony J.},
  title   = {{Macroscopic Quantum Tunneling of a Bose-Einstein Condensate with Attractive Interaction}},
  journal = {Phys. Rev. Lett.},
  volume  = {80}, pages = {1576}, year = {1998},
  doi     = {10.1103/PhysRevLett.80.1576}
}

@article{VakhitovKolokolov1973,
  author  = {Vakhitov, N. G. and Kolokolov, A. A.},
  title   = {{Stationary solutions of the wave equation in a medium with nonlinearity saturation}},
  journal = {Radiophys. Quantum Electron.},
  volume  = {16}, pages = {783}, year = {1973},
  doi     = {10.1007/BF01031343}
}

@article{BaoDu2004,
  author  = {Bao, Weizhu and Du, Qiang},
  title   = {{Computing the ground state solution of Bose-Einstein condensates by a normalized gradient flow}},
  journal = {SIAM J. Sci. Comput.},
  volume  = {25}, pages = {1674}, year = {2004},
  doi     = {10.1137/S1064827503422956}
}

@article{BaoCai2013,
  author  = {Bao, Weizhu and Cai, Yongyong},
  title   = {{Mathematical theory and numerical methods for Bose-Einstein condensation}},
  journal = {Kinet. Relat. Models},
  volume  = {6}, pages = {1}, year = {2013},
  doi     = {10.3934/krm.2013.6.1}
}

@article{Membrado1989,
  author  = {Membrado, M. and Pacheco, A. F. and Sa{\~n}udo, J.},
  title   = {{Hartree solutions for the self-Yukawian boson sphere}},
  journal = {Phys. Rev. A},
  volume  = {39}, pages = {4207}, year = {1989},
  doi     = {10.1103/PhysRevA.39.4207}
}

@article{Lieb1977,
  author  = {Lieb, Elliott H.},
  title   = {{Existence and uniqueness of the minimizing solution of Choquard's nonlinear equation}},
  journal = {Stud. Appl. Math.},
  volume  = {57}, pages = {93}, year = {1977},
  doi     = {10.1002/sapm197757293}
}

@article{LiebSeiringerYngvason2000,
  author  = {Lieb, Elliott H. and Seiringer, Robert and Yngvason, Jakob},
  title   = {{Bosons in a trap: A rigorous derivation of the Gross-Pitaevskii energy functional}},
  journal = {Phys. Rev. A},
  volume  = {61}, pages = {043602}, year = {2000},
  doi     = {10.1103/PhysRevA.61.043602}
}

@article{ElgartSchlein2007,
  author  = {Elgart, Alexander and Schlein, Benjamin},
  title   = {{Mean field dynamics of boson stars}},
  journal = {Commun. Pure Appl. Math.},
  volume  = {60}, pages = {500}, year = {2007},
  doi     = {10.1002/cpa.20134}
}

@article{Coleman1985,
  author  = {Coleman, Sidney},
  title   = {{Q-balls}},
  journal = {Nucl. Phys. B},
  volume  = {262}, pages = {263}, year = {1985},
  doi     = {10.1016/0550-3213(85)90286-X}
}

@article{KleihausKunzList2005,
  author  = {Kleihaus, Burkhard and Kunz, Jutta and List, Meike},
  title   = {{Rotating boson stars and Q-balls}},
  journal = {Phys. Rev. D},
  volume  = {72}, pages = {064002}, year = {2005},
  doi     = {10.1103/PhysRevD.72.064002}
}

@misc{becGPP,
  author       = {Le, Bao-Duy},
  title        = {{becGPP: a Gross-Pitaevskii-Poisson ground-state solver}},
  howpublished = {\url{https://github.com/Duyle0503/becGPP}},
  year         = {2026}
}

@article{CasiniHuertaMyers2011,
  author  = {Casini, Horacio and Huerta, Marina and Myers, Robert C.},
  title   = {{Towards a derivation of holographic entanglement entropy}},
  journal = {J. High Energy Phys.}, volume = {05}, pages = {036}, year = {2011},
  doi     = {10.1007/JHEP05(2011)036}
}

@article{HawkingHunterTaylorRobinson1999,
  author  = {Hawking, S. W. and Hunter, C. J. and Taylor-Robinson, M. M.},
  title   = {{Rotation and the AdS/CFT correspondence}},
  journal = {Phys. Rev. D}, volume = {59}, pages = {064005}, year = {1999},
  doi     = {10.1103/PhysRevD.59.064005}
}

@misc{Ketterle1999,
  author        = {Ketterle, W. and Durfee, D. S. and Stamper-Kurn, D. M.},
  title         = {{Making, probing and understanding Bose-Einstein condensates}},
  year          = {1999},
  eprint        = {cond-mat/9904034},
  archivePrefix = {arXiv},
  note          = {in Proc. Int. School of Physics ``Enrico Fermi'', Course CXL}
}

@book{Helgason1999,
  author    = {Helgason, Sigurdur},
  title     = {{The Radon Transform}},
  edition   = {2nd},
  publisher = {Birkh{\"a}user},
  address   = {Boston},
  year      = {1999}
}

@book{Natterer2001,
  author    = {Natterer, Frank},
  title     = {{The Mathematics of Computerized Tomography}},
  series    = {Classics in Applied Mathematics},
  volume    = {32},
  publisher = {SIAM},
  address   = {Philadelphia},
  year      = {2001}
}

@book{HockneyEastwood1988,
  author    = {Hockney, R. W. and Eastwood, J. W.},
  title     = {{Computer Simulation Using Particles}},
  publisher = {Adam Hilger},
  address   = {Bristol},
  year      = {1988}
}

@article{Kurusa1991,
  author  = {Kurusa, {\'A}rp{\'a}d},
  title   = {{The Radon transform on hyperbolic space}},
  journal = {Geom. Dedicata}, volume = {40}, pages = {325--339}, year = {1991},
  doi     = {10.1007/BF00189917}
}

@article{BerensteinTarabusiKurusa1997,
  author  = {Berenstein, Carlos A. and Casadio Tarabusi, Enrico and Kurusa, {\'A}rp{\'a}d},
  title   = {{Radon transform on spaces of constant curvature}},
  journal = {Proc. Amer. Math. Soc.}, volume = {125}, pages = {455--461}, year = {1997},
  doi     = {10.1090/S0002-9939-97-03570-3}
}

@article{CaoQiSwingleTang2020,
  author  = {Cao, ChunJun and Qi, Xiao-Liang and Swingle, Brian and Tang, Eugene},
  title   = {{Building bulk geometry from the tensor Radon transform}},
  journal = {J. High Energy Phys.}, volume = {2020}, number = {12}, pages = {033}, year = {2020},
  doi     = {10.1007/JHEP12(2020)033},
  eprint  = {2007.00004}
}

@article{Jokela2025,
  author  = {Jokela, Niko and Liimatainen, Tony and Sarkkinen, Miika and Tzou, Leo},
  title   = {{Bulk metric reconstruction from entanglement data via minimal surface area variations}},
  journal = {J. High Energy Phys.},
  volume  = {2025}, number = {10}, pages = {079}, year = {2025},
  doi     = {10.1007/JHEP10(2025)079},
  eprint  = {2504.07016}
}

@article{Gross1961,
  author  = {Gross, Eugene P.},
  title   = {{Structure of a quantized vortex in boson systems}},
  journal = {Nuovo Cimento}, volume = {20}, pages = {454}, year = {1961},
  doi     = {10.1007/BF02731494}
}

@article{Pitaevskii1961,
  author  = {Pitaevskii, Lev P.},
  title   = {{Vortex lines in an imperfect Bose gas}},
  journal = {Sov. Phys. JETP}, volume = {13}, pages = {451}, year = {1961}
}

@article{LiebThirring1984,
  author  = {Lieb, Elliott H. and Thirring, Walter E.},
  title   = {{Gravitational collapse in quantum mechanics with relativistic kinetic energy}},
  journal = {Ann. Phys. (N.Y.)}, volume = {155}, pages = {494}, year = {1984},
  doi     = {10.1016/0003-4916(84)90010-1}
}

@article{LewinNamRougerie2014,
  author  = {Lewin, Mathieu and Nam, Phan Th{\`a}nh and Rougerie, Nicolas},
  title   = {{Derivation of Hartree's theory for generic mean-field Bose systems}},
  journal = {Adv. Math.}, volume = {254}, pages = {570}, year = {2014},
  doi     = {10.1016/j.aim.2013.12.010}
}

@article{GiuliniGrossardt2012,
  author  = {Giulini, Domenico and Gro{\ss}ardt, Andr{\'e}},
  title   = {{The Schr\"odinger-Newton equation as a non-relativistic limit of self-gravitating Klein-Gordon and Dirac fields}},
  journal = {Class. Quantum Grav.}, volume = {29}, pages = {215010}, year = {2012},
  doi     = {10.1088/0264-9381/29/21/215010}
}

@article{IyerWald1994,
  author  = {Iyer, Vivek and Wald, Robert M.},
  title   = {{Some properties of the Noether charge and a proposal for dynamical black hole entropy}},
  journal = {Phys. Rev. D},
  volume  = {50}, pages = {846}, year = {1994},
  doi     = {10.1103/PhysRevD.50.846}
}

@misc{SwingleVanRaamsdonk2014,
  author        = {Swingle, Brian and Van Raamsdonk, Mark},
  title         = {{Universality of gravity from entanglement}},
  year          = {2014},
  eprint        = {1405.2933},
  archivePrefix = {arXiv},
  primaryClass  = {hep-th}
}

@article{deBoerHellerMyersNeiman2016,
  author  = {de Boer, Jan and Heller, Michal P. and Myers, Robert C. and Neiman, Yasha},
  title   = {{Holographic de Sitter geometry from entanglement in conformal field theory}},
  journal = {Phys. Rev. Lett.}, volume = {116}, pages = {061602}, year = {2016},
  doi     = {10.1103/PhysRevLett.116.061602}, eprint = {1509.00113}
}

@article{deBoerHaehlHellerMyers2016,
  author  = {de Boer, Jan and Haehl, Felix M. and Heller, Michal P. and Myers, Robert C.},
  title   = {{Entanglement, holography and causal diamonds}},
  journal = {J. High Energy Phys.}, volume = {2016}, number = {8}, pages = {162}, year = {2016},
  doi     = {10.1007/JHEP08(2016)162}, eprint = {1606.03307}
}

@article{CzechLamprouMcCandlishSully2015,
  author  = {Czech, Bart{\l}omiej and Lamprou, Lampros and McCandlish, Samuel and Sully, James},
  title   = {{Integral geometry and holography}},
  journal = {J. High Energy Phys.}, volume = {2015}, number = {10}, pages = {175}, year = {2015},
  doi     = {10.1007/JHEP10(2015)175}, eprint = {1505.05515}
}

@article{CzechEtAl2017,
  author  = {Czech, Bart{\l}omiej and Lamprou, Lampros and McCandlish, Samuel and Mosk, Benjamin and Sully, James},
  title   = {{Equivalent equations of motion for gravity and entropy}},
  journal = {J. High Energy Phys.}, volume = {2017}, number = {2}, pages = {004}, year = {2017},
  doi     = {10.1007/JHEP02(2017)004}, eprint = {1608.06282}
}

@article{Hubeny2012,
  author  = {Hubeny, Veronika E.},
  title   = {{Extremal surfaces as bulk probes in AdS/CFT}},
  journal = {J. High Energy Phys.}, volume = {2012}, number = {7}, pages = {093}, year = {2012},
  doi     = {10.1007/JHEP07(2012)093}, eprint = {1203.1044}
}

@misc{Nogueira2013,
  author        = {Nogueira, Fernando},
  title         = {{Extremal surfaces in asymptotically AdS charged boson stars backgrounds}},
  year          = {2013},
  eprint        = {1301.4316},
  archivePrefix = {arXiv},
  primaryClass  = {hep-th}
}

@article{Freivogel2015,
  author  = {Freivogel, Ben and Jefferson, Robert A. and Kabir, Laurens and Mosk, Benjamin and Yang, I-Sheng},
  title   = {{Casting shadows on holographic reconstruction}},
  journal = {Phys. Rev. D}, volume = {91}, pages = {086013}, year = {2015},
  doi     = {10.1103/PhysRevD.91.086013}, eprint = {1412.5175}
}

@article{DiasHorowitzSantos2011,
  author  = {Dias, {\'O}scar J. C. and Horowitz, Gary T. and Santos, Jorge E.},
  title   = {{Black holes with only one Killing field}},
  journal = {J. High Energy Phys.}, volume = {2011}, number = {7}, pages = {115}, year = {2011},
  doi     = {10.1007/JHEP07(2011)115}, eprint = {1105.4167}
}

\end{document}